\documentclass{aastex}
\usepackage{emulateapj5,times,mathptm}

\newcommand  \kms      {\ifmmode {\rm km\,s}^{-1} \else km\,s$^{-1}$\fi}

\newcommand  \cmii     {\hbox{cm$^{-2}$}}
\newcommand  \ergs     {\ifmmode {\rm ergs\,s}^{-1} \else ergs s$^{-1}$\fi}
\newcommand  \ergcms   {\ifmmode {\rm ergs\,cm}^{-2}\,{\rm s}^{-1}
                        \else ergs\,cm$^{-2}$\,s$^{-1}$\fi}
\newcommand  \ergcmsA  {\ifmmode{\rm ergs\,cm}^{-2}\,{\rm s}^{-1}\,{\rm\AA}^{-1}
                        \else ergs\,cm$^{-2}$\,s$^{-1}$\,\AA$^{-1}$\fi}
\newcommand  \ergcmsHz {\ifmmode{\rm ergs\,cm}^{-2}\,{\rm s}^{-1}\,{\rm Hz}^{-1}
                        \else ergs\,cm$^{-2}$\,s$^{-1}$\,Hz$^{-1}$\fi}
\newcommand  \phcms    {\ifmmode {\rm ph\,cm}^{-2}\,{\rm s}^{-1}
                        \else ,ph\,cm$^{-2}$\,s$^{-1}$\fi}
\newcommand  \phcmsA   {\ifmmode {\rm ph\,cm}^{-2}\,{\rm s}^{-1}\,{\rm\AA}^{-1}
                        \else ph\,cm$^{-2}$\,s$^{-1}$\,\AA$^{-1}$\fi}

\journalinfo{The Astrophysical Journal Supplements Series, 
14?: ??1--?19, 2002 ???, astro-ph/0203263}
\slugcomment{Received 2002 January 17; accepted 2002 March 13}

\shorttitle{900 KS {\it CHANDRA} GRATING SPECTROSCOPY OF NGC\,3783}

\shortauthors{KASPI ET AL.}

\begin{document}

\title{The Ionized Gas and Nuclear Environment in NGC\,3783. \\
I. Time-Averaged 900 ks {\it Chandra} Grating Spectroscopy}
\author{
Shai Kaspi,\altaffilmark{1,2} 
W. N. Brandt,\altaffilmark{1} 
Ian M. George,\altaffilmark{3,4}
Hagai Netzer,\altaffilmark{2}
D. Michael Crenshaw,\altaffilmark{5}
Jack R. Gabel,\altaffilmark{6}
Frederick W. Hamann,\altaffilmark{7}
Mary Elizabeth Kaiser,\altaffilmark{8}
Anuradha Koratkar,\altaffilmark{9}
Steven B. Kraemer,\altaffilmark{6}
Gerard A. Kriss,\altaffilmark{8,9}
Smita Mathur,\altaffilmark{10}
Richard F. Mushotzky,\altaffilmark{3}
Kirpal Nandra,\altaffilmark{3,11}
Bradley M. Peterson,\altaffilmark{10}
Joseph C. Shields,\altaffilmark{12}
T. J. Turner,\altaffilmark{3,4}
and
Wei Zheng\altaffilmark{8}
}

\altaffiltext{1}{Department of Astronomy and Astrophysics, 525 Davey
Laboratory, The Pennsylvania State University, University Park, PA 16802.}
\altaffiltext{2}{School of Physics and Astronomy, Raymond and Beverly Sackler
Faculty of Exact Sciences, Tel-Aviv University, Tel-Aviv 69978, Israel.}
\altaffiltext{3}{Laboratory for High Energy Astrophysics, NASA/Goddard Space
Flight Center, Code 662, Greenbelt, MD 20771.}
\altaffiltext{4}{Joint Center for Astrophysics, Physics Department, University
of Maryland, Baltimore County, 1000 Hilltop Circle, Baltimore, MD 21250.}
\altaffiltext{5}{Department of Physics and Astronomy, Georgia State
University, Atlanta, GA 30303.}
\altaffiltext{6}{Catholic University of America, NASA/GSFC, Code 681,
Greenbelt, MD 20771.} 
\altaffiltext{7}{Department of Astronomy, University of Florida, 211 Bryant 
Space Science Center, Gainesville, FL 32611-2055.}
\altaffiltext{8}{Center for Astrophysical Sciences, Department of Physics and
Astronomy, The Johns Hopkins University, Baltimore, MD 21218-2686.}
\altaffiltext{9}{Space Telescope Science Institute, 3700 San Martin Drive,
Baltimore, MD 21218.}
\altaffiltext{10}{Department of Astronomy, Ohio State University, 140 West 18th
Avenue, Columbus, OH 43210-1106.}
\altaffiltext{11}{Universities Space Research Association, 7501 Forbes Boulevard,
Suite 206, Seabrook, MD 207006-2253.} 
\altaffiltext{12}{Department of Physics and Astronomy, Clippinger
Research Labs 251B, Ohio University, Athens, OH 45701-2979.}

\begin{abstract}
We present results from a 900 ks exposure of NGC\,3783 with the
High-Energy Transmission Grating Spectrometer on board the {\it Chandra
X-ray Observatory\/}. The resulting X-ray spectrum, which covers the
0.5--10 keV energy range, has the best combination of signal-to-noise
and resolution ever obtained for an AGN. This spectrum reveals
absorption lines from H-like and He-like ions of N, O, Ne, Mg, Al, Si,
and S. There are also possible absorption lines from H-like and He-like
Ar and Ca as well as H-like C. We also identify inner-shell absorption
from lower-ionization ions such as \ion{Si}{7}--\ion{Si}{12} and
\ion{S}{12}--\ion{S}{14}.  The iron absorption spectrum is very rich;
L-shell lines of \ion{Fe}{17}--\ion{Fe}{24} are detected, as well as
probable resonance lines from \ion{Fe}{25}. A strong complex of M-shell
lines from iron ions is also detected in the spectrum The absorption
lines are blueshifted relative to the systemic velocity by a mean
velocity of $-590\pm150$~\kms. We resolve many of the absorption lines,
and their mean FWHM is $820\pm280$~\kms. We do not find correlations
between the velocity shifts or the FWHMs with the ionization potentials
of the ions. Most absorption lines show asymmetry, having more
extended blue wings than red wings. In \ion{O}{7} we have resolved
this asymmetry to be from an additional absorption system at $\sim
-1300$~\kms. The two X-ray absorption systems are consistent in
velocity shift and FWHM with the ones identified in the UV lines of
\ion{C}{4}, \ion{N}{5}, and \ion{H}{1}. Equivalent width measurements
for all absorption and emission lines are given and column densities
are calculated for several ions. We resolve the narrow Fe\,K$\alpha$
line at $6398.2\pm3.3$ eV to have a FWHM of $1720\pm360$~\kms, which
suggests that this narrow line may be emitted from the outer part of
the broad line region or the inner part of the torus. We also detect a
``Compton shoulder'' redward of the narrow Fe\,K$\alpha$ line which
indicates that it arises in cold, Compton-thick gas.
\end{abstract}


\section{Introduction}

NGC\,3783 is one of the best-studied and brightest Seyfert~1 galaxies
($V\approx 13.5$ mag). It has some of the strongest X-ray absorption
features around 0.7--1.5 keV known for a Seyfert~1; these have
been typically attributed to \ion{O}{7} (739 eV) and \ion{O}{8}
(871 eV) edges indicating the presence of a ``warm absorber.'' NGC\,3783
has been studied in detail in the X-ray band with {\em ROSAT}
(Turner et al. 1993) and {\em ASCA} (e.g., George et al. 1998) as
well as with the new generation of X-ray observatories, {\it Chandra}
(Kaspi et al. 2000a, 2001) and {\em XMM-Newton} (J. Blustin et
al., in preparation). Its 2--10~keV spectrum is fitted by a power
law with photon index $\Gamma \approx 1.7$--1.8, the 2--10 keV flux
varies in the range $\sim(4$--$9)\times 10^{-11}$~\ergcms , and
its mean X-ray luminosity is $\sim 1.5\times 10^{43}$~\ergs\ (for
$H_0=70$~km\,s$^{-1}$\,Mpc$^{-1}$ and $q_0=0.5$). Modeling the apparent
\ion{O}{7} and \ion{O}{8} edges indicates a column density of ionized
gas of $\sim 2 \times 10^{22}$~cm$^{-2}$.

\begin{table*}[t]
\footnotesize    
\begin{center}
\caption{{\it Chandra} Observation Log of NGC\,3783 \label{obslog}}
\begin{tabular}{ccccc}
\hline
\hline
{Sequence Number} &
{UT start} &
{UT end} &
{Time (ks)\tablenotemark{a}} &
{Roll angle ($\degr$)\tablenotemark{b}} \\
\hline
700045 & 2000 Jan 20, 23:33 & 2000 Jan 21, 16:20 & {\phn}56.4 & {\phn}50.1 \\
700280 & 2001 Feb 24, 18:45 & 2001 Feb 26, 17:48 & 165.7 &  {\phn}22.1 \\
700281 & 2001 Feb 27, 09:18 & 2001 Mar 01, 09:10 & 168.8 &  {\phn}19.4 \\
700282 & 2001 Mar 10, 00:31 & 2001 Mar 11, 23:30 & 165.5 &  {\phn\phn}6.7  \\
700283 & 2001 Mar 31, 03:36 & 2001 Apr 02, 02:48 & 166.1 &  334.5 \\
700284 & 2001 Jun 26, 09:57 & 2001 Jun 28, 09:10 & 166.2 &  245.4 \\
\hline
\end{tabular}
\vskip 2pt
\parbox{4.2in}{ 
\small\baselineskip 9pt
\footnotesize
\indent
$\rm ^a${Sum of good time intervals corrected for detector
dead time (LIVETIME).} \\
$\rm ^b${Roll angle describes the orientation of the {\it
Chandra} instruments on the sky. The angle increases to the West of
North --- opposite to the traditional position angle.}
}
\end{center}
\end{table*}
\normalsize

NGC\,3783 was observed for 56~ks with the High-Energy Transmission
Grating Spectrometer (HETGS; C. R. Canizares et al., in preparation)
on the {\it Chandra X-ray Observatory\/}\footnote{See {\it The Chandra
Proposers' Observatory Guide} at http://asc.harvard.edu/udocs/docs/.}
with the Advanced CCD Imaging Spectrometer (ACIS; G. P. Garmire et
al., in preparation) as the detector. The observation and resulting
high-resolution spectrum (covering the 1.6--23.4\,\AA\ wavelength range) are described by Kaspi et al. (2000a, 2001).
This spectrum shows several dozen absorption lines and a few emission
lines from the H-like and He-like ions of O, Ne, Mg, Si, and S as well as
from \ion{Fe}{17}--\ion{Fe}{23} L-shell transitions. The absorption lines
are blueshifted relative to the systemic velocity by $\sim -610$~\kms\
while the emission lines are consistent with being at the systemic
velocity (throughout this paper we use a redshift of $0.009760\pm
0.000093$ which corresponds to a systemic velocity of $2926\pm28$~\kms;
de Vaucouleurs et al. 1991). With the limited signal-to-noise ratio (S/N)
of this observation, only the stacked composite of several Ne absorption
lines could be resolved to have a FWHM of $840^{+490}_{-360}$~\kms . The
high-resolution X-ray spectrum was modeled with two absorption components,
with different global covering factors and an order of magnitude
difference in their ionization parameters. The two components were taken
to be spherical shells of highly ionized gas radially outflowing from
the AGN and thus contribute to both the absorption and the emission
via P Cygni profiles. A two, or more, absorption-component model was
also suggested by George et al. (1998) to explain the warm-absorber
variability seen by {\it ASCA}.

The UV spectrum of NGC\,3783 shows intrinsic absorption features due to
\ion{C}{4}, \ion{N}{5}, and \ion{H}{1} (e.g., Kraemer, Crenshaw, \& Gabel
2001 and references therein). Currently there are three known absorption
systems in the UV at radial velocities of $\sim -$560, $-$720,
and $-$1400~\kms\ (blueshifted) relative to the optical redshift,
with FWHMs of 170, 280, and 190 \kms, respectively. The
strength of the absorption is found to be variable over time scales of
months to years. The relation between the X-ray and the UV absorbers is
not clear. It has been suggested by several studies that both types of
absorption might arise from the same gas component in the AGN (e.g.,
Mathur, Elvis, \& Wilkes 1995; Shields \& Hamann 1997), but no firm
conclusions could be reached. The main unknowns were the spectral energy
distribution (SED) at far-UV energies, responsible for the ionization of
the observed UV species, and the poor X-ray spectral resolution. With
the better resolution of the {\it Chandra}/HETGS, Kaspi et al. (2001)
found the absorbing X-ray system to have an outflow velocity that is
consistent with the two smallest outflow velocities of the UV absorbing
systems. However, no X-ray absorbing system was found to correspond to the
UV absorbing system at $-$1400~\kms . We note that Kraemer et al. (2001)
predict \ion{O}{7} and \ion{O}{8} X-ray absorption lines based on the
modeling of the UV absorption at $-$1400~\kms. The low S/N of the 56 ks
observation might have prevented detection of this absorption system in
the X-ray band.

We selected NGC\,3783 for a long multiwavelength monitoring program to
study in detail its nuclear spectrum over time and to address further
the above issues. NGC\,3783 is most suitable for such a program as
it is very bright in the X-ray and UV, has one of the strongest warm
absorbers, has shown narrow absorption lines both in the X-ray and the
UV bands, and has shown spectral variability over time in both bands.
The monitoring campaign, involving {\it Chandra}/HETGS and {\em RXTE}
in the X-ray, {\em HST}/STIS and {\em FUSE} in the UV, and ground-based
observations, took place during 2001 February--June. The {\it HST}/STIS
and {\em FUSE} observations as well as the time variability of the
X-ray data are described in associated papers (D. M. Crenshaw et al.,
in preparation; J. R. Gabel et al., in preparation; I. M. George et
al., in preparation). Detailed photoionization modeling of the X-ray
and UV spectra will be discussed in subsequent papers. In this paper
we focus on the average high-resolution X-ray spectrum of NGC\,3783
obtained from combining 900 ks of exposure time. In \S2 we describe
the observations and reduction which result in the 900 ks X-ray
spectrum. In \S3 we study the absorption lines by resolving them and
measuring their dynamical properties. In \S4 we discuss the emission
lines, in \S5 we study the spectrum from 4--9 keV focusing on the
Fe\,K$\alpha$ line, and in \S6 we summarize our results.

\section{Observations and Data Reduction}

Five observations of NGC\,3783 (each of $\sim 170$~ks duration)
were taken with the {\em Chandra}/HETGS during the period 2001
February--June. A log of the new observations together with the
observation from Kaspi et al. (2001) is given in Table~\ref{obslog}. The
total exposure time (ONTIME) of the six observations is 900.1 ks, and
the total good time interval corrected for detector dead time (LIVETIME)
is 888.7~ks. All observations were reduced uniformly and in the standard
way using the {\it Chandra} Interactive Analysis of Observations (CIAO)
software (Version 2.1.2) and its threads of 2001 July 3. Flux calibration
of each observation was carried out using the {\em Chandra} Calibration
Database (Version~2.6). All spectra described below were also corrected
for Galactic absorption ($N_{\rm H} = 8.7 \times 10^{20}$~\cmii; Alloin et
al. 1995), and the wavelengths were redshift corrected and are presented
in the rest frame of NGC\,3783.

From the combined zeroth-order images we find the X-ray centroid
of NGC\,3783 to be at $\alpha_{2000}=$~11$^{\rm h}$39$^{\rm
m}$01$\fs$7, $\delta_{2000}=$~$-37\arcdeg$44$\arcmin$19$\farcs$0
with accuracy of $\sim$0$\farcs$5. This is identical to the
X-ray position reported in Kaspi et al. (2001), and consistent
with the nuclear position in the radio reported by Ulvestad \&
Wilson (1984; $\alpha_{2000}=$~11$^{\rm h}$39$^{\rm m}$01$\fs$72,
$\delta_{2000}=$~$-37\arcdeg$44$\arcmin$19$\farcs$3). As described
further in I. M. George et al. (in preparation), we detect a number
of serendipitous sources surrounding the nucleus, but these and any
extended X-ray emission are at such a low intensity that they do not
affect any of the conclusions presented here.

The HETGS produces high-order spectra from two grating assemblies,
the medium-energy grating (MEG) and the high-energy grating (HEG).
Both positive and negative orders are imaged by the ACIS-S array.
We combined the $+$1st and $-$1st orders for each of the MEG and HEG
spectra by averaging them using a 1/$\sigma^2$ weighted mean (where
$\sigma$ is the uncertainty) to produce mean MEG and HEG spectra for
each observation. We checked for variability within the individual
observations and between different observations. Variability is found
(at a level of about 30\% around the median within the observations
and about 50\% around the median between the observations) and will be
discussed in detail in I.M. George et al. (in preparation). This variability
does not materially affect the key results reported here. We have
verified that the ACIS chip gaps do not produce any false spectral
features in the spectra by detailed comparison of the HETGS response
curve of the observation (the ARF) to the final spectrum. The chip
gaps are broader than the features discussed here and do not seem
to produce any features in the flux calibrated spectrum. Furthermore,
our averaging method minimizes any possible effect of the chip gaps
as explained in \S2 of Kaspi et al. (2001).

In the following sections we have further combined different subsets of
the spectra together. For example, all six MEG spectra and all six HEG
spectra were combined together to produce 900~ks MEG and HEG spectra,
respectively. In all cases here and hereafter the spectra were combined
using the 1/$\sigma^2$ weighted-mean method. The total numbers of counts
in the first order in the energy range 0.5--10 keV are 583,196 and
313,861 for the MEG and HEG, respectively. The S/Ns at $\sim 7$~\AA\
of the combined 900 ks spectra are $\sim 23$ and $\sim 10$ for the MEG
and HEG, respectively, when using the default CIAO bins of 0.005~\AA\
for the MEG and 0.0025~\AA\ for the HEG (compare the S/Ns of $\sim 5$
and $\sim 2.5$ reported in Kaspi et al. 2001).

The details of the 56 ks observation are discussed in Kaspi et al.
(2001). The five new observations were all continuous without any
noticeable time gaps or special events during the observations. We
checked the background counts for flares and found none. The background
level of the HETGS/ACIS is $\la 0.5$\% of the signal which is negligible;
hence we did not subtract any background in the following analysis.

We also extracted spectra from the second and third orders for our
individual observations and combined them into average spectra in the
same way as described above. The resolutions of the second and third
order spectra are better than that of the first order by factors of two
and three, respectively. These higher order spectra agree well with the
spectra extracted from the first order. However, since the effective area
of these orders is about an order of magnitude less than the effective
area of the first order, the utility of the higher order spectra is
limited. The highest effective area (after the first orders) is for the
MEG third order (resolution of 0.0077 \AA ). In the 900 ks MEG third-order
spectrum the S/N is $\sim 2$ when using a bin size of 0.0033~\AA .
In addition, the wavelength range of the second and third orders is
smaller than that of the first order by factors of two and three,
respectively; this further limits the utility of the higher order spectra.

\begin{figure*}
\centerline{\includegraphics[width=18cm]{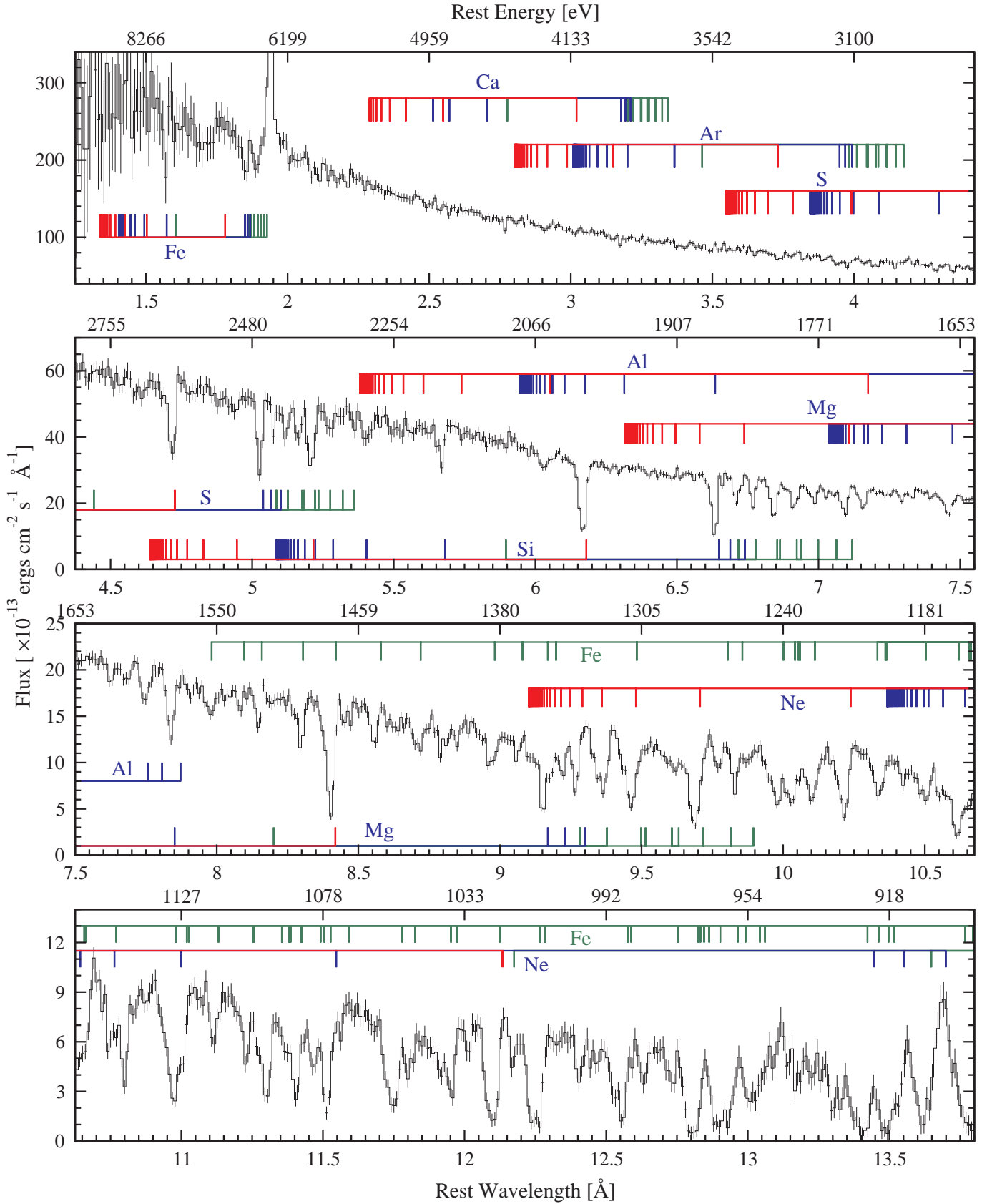}}
\caption{Combined MEG and HEG first-order 900 ks spectrum
binned to 0.01 \AA . Each data point has an error bar representing its
1$\sigma$ uncertainty. The H-like and He-like lines of the identified
ions are marked in red and blue, respectively. Lines from other ions
(lower ionization metals and \ion{Fe}{17} to \ion{Fe}{24}) are marked
in green. For each H-like or He-like ion the theoretically expected
lines are plotted up to the ion's edge (not all lines are identified in
the data). The ions' lines are marked at their expected wavelengths in
the rest frame of NGC\,3783, and the blueshift of the absorption lines
is noticeable.
\label{xrayspec} }
\end{figure*}

\setcounter{figure}{0}

\begin{figure*}
\centerline{\includegraphics[width=18cm]{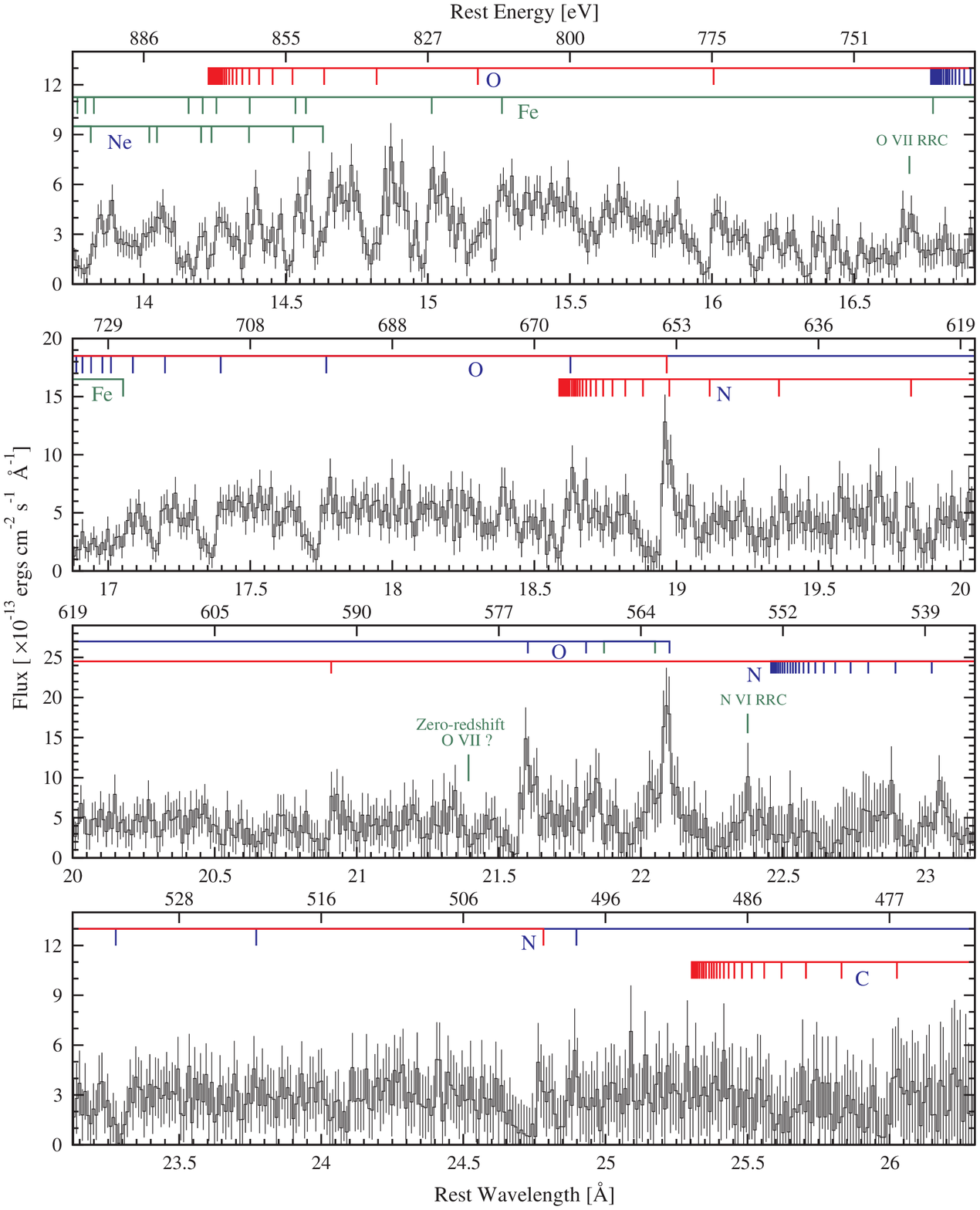}}
\caption{{\it Continued}
\label{xrayspec_b} }
\end{figure*}

\begin{figure*}
\centerline{\includegraphics[width=18.5cm]{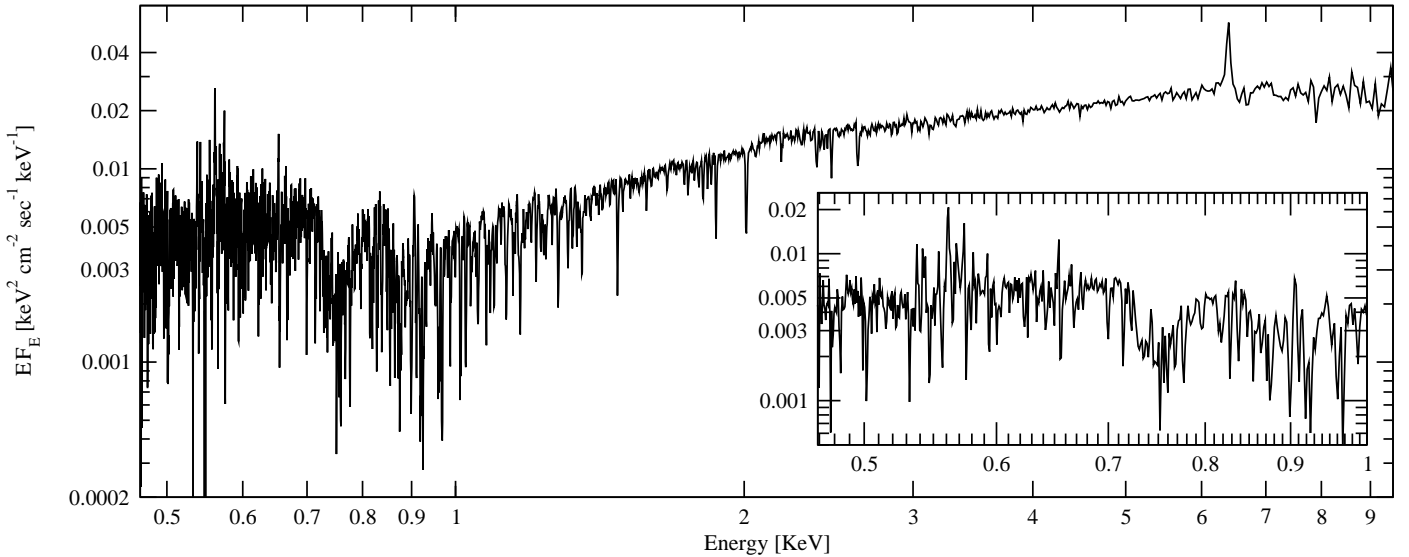}}
\caption{The combined MEG and HEG first-order 900 ks spectrum which
was presented in Figure~\protect\ref{xrayspec} is shown here over the
whole range in $EF\,\,{_{\rm E}}$ vs. energy format. Uncertainties are
not shown for clarity. The warm absorber features are noticeable. The
shape of this spectrum is consistent with previously published {\it ASCA}
spectra (e.g., George et al. 1998). The Insert focuses on the spectrum
below 1 keV binned to 0.03 \AA. No neutral oxygen edge is seen at 538 eV.
\label{continuum} }
\end{figure*}

\section{Absorption Lines}

\subsection{Basic Line Properties}
\label{basic_line}

\begin{figure*}
\centerline{\includegraphics[width=17.2cm]{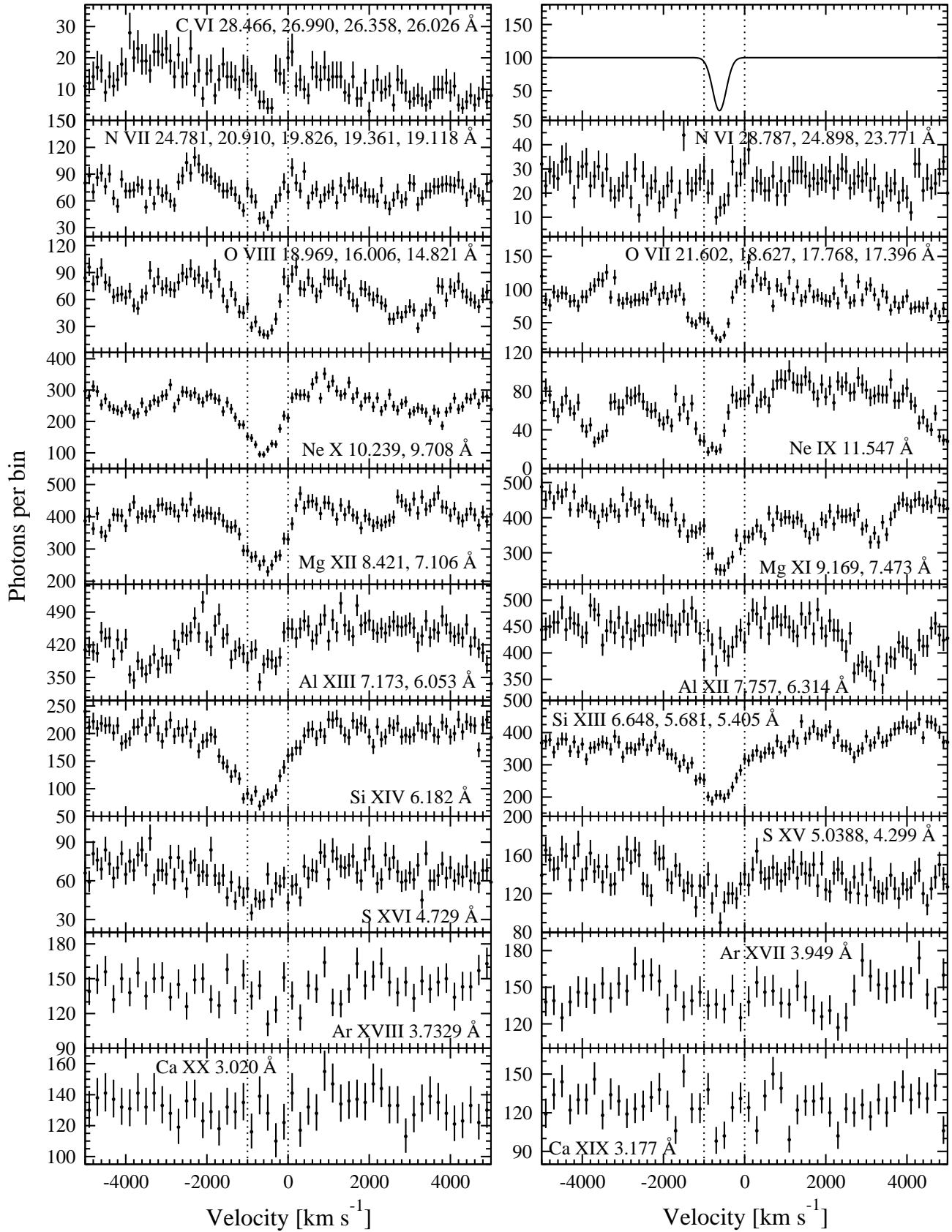}}
\caption{MEG velocity spectra showing co-added lines from
different species. H-like ions are shown on the left, and He-like ions
are shown on the right. The bin size is 100 \kms\ for the upper panels
and 200~\kms\ for the bottom two panels (Ar and Ca). Error bars have been
computed following Gehrels (1986). Absorption is clearly detected for
all abundant elements from N to S. There are hints of absorption by C,
Ar, and Ca, although these are not significant. Vertical dotted lines
are marked at velocities of 0 and $-1000$~\kms\ to guide the eye. In the
uppermost right panel we show a Gaussian absorption line representing
the line response function of the MEG at 17.396~\AA\ (the Gaussian FWHM
is 397~\kms); this is the poorest line response function applicable to
the coadded velocity spectrum of \protect\ion{O}{7}. Note the asymmetry
of the \protect\ion{O}{7} lines that is apparently from an additional
absorption system.
\label{velspec_meg} }
\end{figure*}

\begin{figure*}[t]
\centerline{\includegraphics[width=17.2cm]{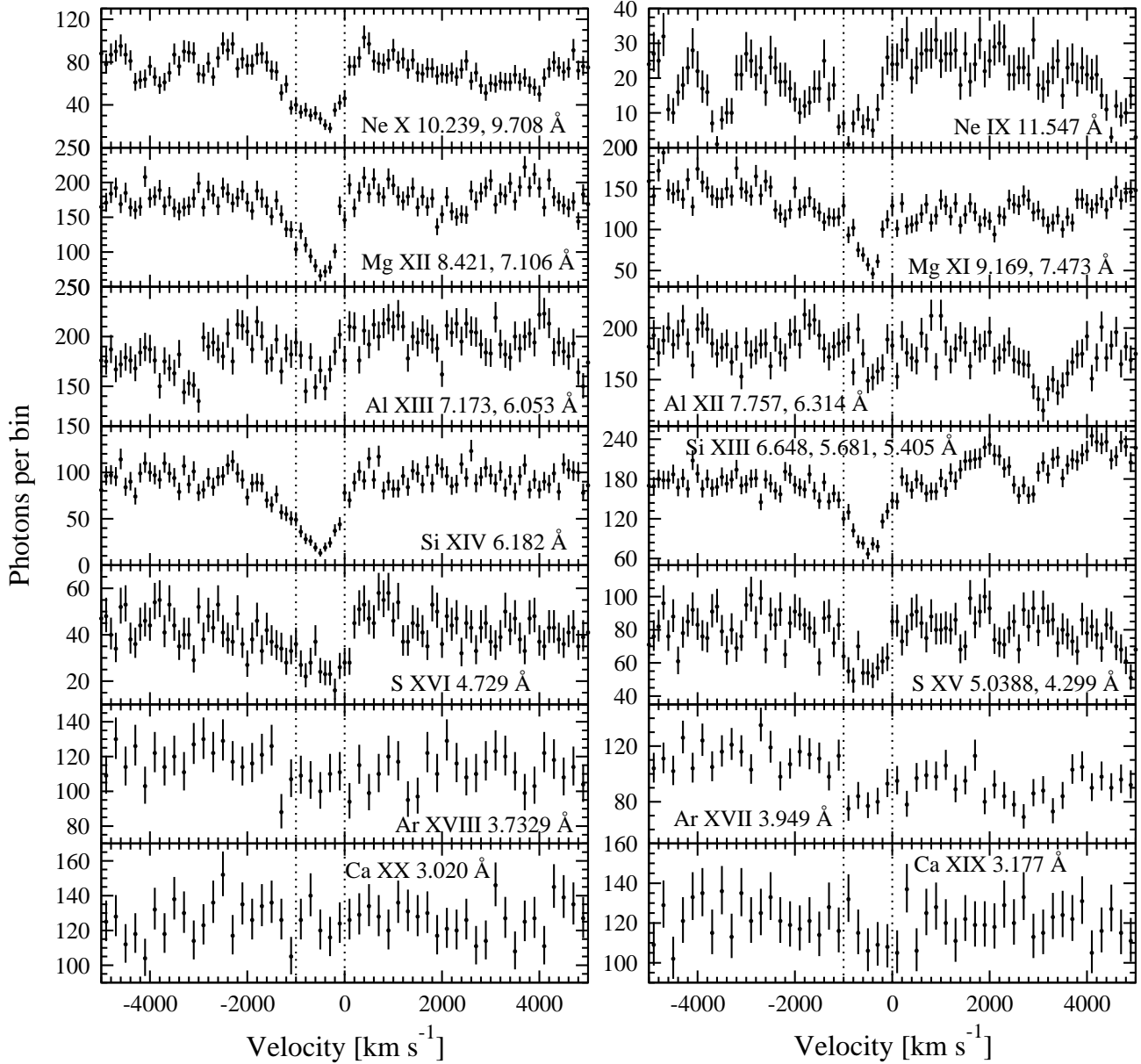}}
\caption{Same as in Figure~\ref{velspec_meg} but showing HEG velocity
spectra. The lines from C, N, and O are not shown since they are not
covered by the HEG energy range.
\label{velspec_heg} }
\end{figure*}

The combined 900 ks spectrum reveals absorption lines from H-like and
He-like ions of N, O, Ne, Mg, Al, Si, and S (see Figure~\ref{xrayspec}).
For several of these elements we also identify lines from less-ionized
ions (see below). There are also many L-shell and M-shell lines of
\ion{Fe}{17}--\ion{Fe}{24} as well as probable resonance lines of
\ion{Fe}{25} (see \S\ref{ironsec}). Figure~\ref{continuum} presents
the entire energy range of this spectrum in one panel to illustrate the
overall continuum, and Figure~\ref{thirdorder} demonstrates the strongest
absorption lines resolved by the MEG third-order. To examine further the
absorption we created ``velocity spectra'' by adding, in velocity space,
several absorption lines from the same ion (see also Kaspi et al. 2001).
The lines were chosen to be the strongest predicted features from a
given ion and to be free from contamination by adjacent features. The
velocity spectra were built up on a photon-by-photon basis (rather than
by interpolating spectra already binned in wavelength). The velocity
spectra are shown in Figures~\ref{velspec_meg} and \ref{velspec_heg}.
Absorption from H-like and He-like ions is shown separately to enable
profile comparison. Absorption is clearly detected for N, O, Ne, Mg,
Al, Si, and S. There are also hints of absorption by C, Ar, and Ca,
although these are not statistically significant.

About 135 absorption-line features can be identified in the spectrum;
many of these are blends of several lines. Since the 900 ks spectrum
spans over an order of magnitude in energy, resolution, and S/N, we
used different combinations and binnings of the spectrum for measuring
the wavelengths and FWHMs of the lines in different wavelength bands. In
most of the measurements, we used the HEG spectrum binned to 0.005 \AA\
at short wavelengths ($\la 12$\AA), and the combined MEG+HEG spectrum
binned to 0.01 \AA\ at longer wavelengths. We fitted a Gaussian to
each line combined with a local continuum. The local continuum was
determined using wavelength bands where no line features are present or
expected (i.e., the ``line free zones'' of Kaspi et al. 2001). We fitted
small regions of the spectrum with a third order polynomial. Since the
segments were small ($\la 2.5$~\AA ) the fitted continuum was close to
a straight line. We note that the real continuum of the source could
be different from the local continuum we determine here hence affecting
the EW measurements below.

The wavelengths and FWHMs of the lines were determined from the Gaussian
fits. However, many line features are blended (mainly with Fe lines in
the 8--17 \AA\ band), and in many cases these measurements represent
the entire blend and not individual lines. The equivalent width (EW)
for each feature was measured directly from the spectrum and not using
the Gaussian fitting. Thus
\begin{equation}
{\rm EW} = \sum_{i}^{}{\left(1-\frac{F_i}{F_c}\right)B_i}
\end{equation}
where $i$ runs over all bins, $F_i$ is the flux in the $i$th bin, $F_c$
is the underlying continuum flux, and $B_i$ is the bin width in \AA.
The EW's uncertainty was derived by propagating the uncertainty on the
flux in each bin and the uncertainty in the continuum placement, i.e.,
\begin{equation}
\Delta{\rm EW} = \sqrt{ \left[\frac{\Delta{F_c}}{F_c}\,\sum_{i}^{}{\left(\frac{B_i F_i}{F_c}\right)}\right]^2
+ \sum_{i}^{}{\left(\frac{B_i\,\Delta{F_i}}{F_c}\right)^2}}
\end{equation}
where $\Delta{F_c}$ is the uncertainty in the mean continuum flux
in the ``line free zones'' adjacent to the measured feature,
and $\Delta{F_i}$ is the uncertainty in the flux of each bin.
The summation limits were chosen at the endpoints of the line feature
where the profile starts to become displaced from the continuum.
The EW uncertainties can be used as a measure to assess the detection
of a line. We consider lines with EW measurements higher than 2$\sigma$
to be clear detection, while the detections of lines with EW values
less than that should be considered with caution.  Only a few simple
cases allowed line deblending, while other blends are more complex
and were measured as such. The line measurements are presented in
Table~\ref{linetab}. For each line we list the 
%
\centerline{\includegraphics[width=8.5cm]{f3.eps}}
\figcaption{The three strongest individual absorption lines from the MEG
third-order 900 ks spectrum binned to 0.0033 \AA . The resolution is
0.0077 \AA. All three lines are clearly resolved.
\label{thirdorder} }
%
\noindent central wavelength and
FWHM as determined from the Gaussian fit, the EW and flux derived for
this line (negative EW indicates emission and positive EW indicates
absorption), and the main ions contributing to this feature.

We are able to identify almost all line features with EW$\ga 5$~m\AA
. The exceptions are two lines with EW$\sim$10~m\AA\ at 12.632~\AA\ and
12.404~\AA, one line with EW$\sim$5~m\AA\ at 12.169~\AA, and several
lines with EW$\la$2.5~m\AA\ below 5~\AA\ which are weak and are not
necessarily clear detections. We identified many inner-shell lines
of \ion{S}{14}, \ion{S}{13}, \ion{S}{12}, \ion{Si}{12}, \ion{Si}{11},
\ion{Si}{10}, \ion{Si}{9}, \ion{Si}{8}, \ion{Si}{7}, \ion{Mg}{10}, and
\ion{Mg}{9}. These are mostly 1$s$--2$p$ inner-shell transitions that
are expected to have comparable oscillator strengths to the identified
H-like and He-like transitions. Atomic data for these lines have recently
been calculated by Behar \& Netzer (2002) and our measurements provide
confirmation for their presence. The detection of these lines provides
valuable information about a wide range of ionization stages for several
elements, and the lines will be used in a future publication to constrain
photoionization models of the absorbing gas.

In order to constrain the contribution of scattered X-ray emission
to our spectrum, we have searched for lines that are ``black'' (i.e.,
that drop to zero intensity) or nearly black. We consider only electron
scattering which has a cross section that is essentially constant
across the {\it Chandra} bandpass. The best such region is near 12.8~\AA\
with a blend of two strong \ion{Fe}{20} lines (12.846 and 12.864 \AA;
see Figure~\ref{xrayspec}). At the bottom of this feature the flux is
measured to be $(0.57\pm0.20)\times 10^{-13}$ \ergcmsA\ while the local
continuum flux is much larger ($[5.35\pm0.47]\times 10^{-13}$ \ergcmsA
). This suggests that there is little or no scattering of X-rays around
the absorption (i.e., the line-of-sight covering fraction is $\sim
1$); formally we limit the scattered contribution to be $\la 15$\%
[using the \ion{Fe}{20} (12.846, 12.864) lines].

\subsection{Special Absorption Features}

A broad absorption feature is seen in the spectrum around 15.5--17~\AA\
(Figure~\ref{uta}). \ We \ identify this feature as arising from 

\noindent numerous
inner-shell 2$p$--3$d$ absorption lines of Fe M-shell ions blended into
an unresolved transition array (UTA; Sako et al. 2001; Behar, Sako, \&
Kahn 2001). This absorption feature is blended with the \ion{O}{7}
(16.778 \AA) edge making it impossible to measure the relative
intensity of the two.  Furthermore, the UTA is also blended with the
many \ion{O}{7} absorption lines close to the edge. The presence of
the \ion{O}{7} edge, as well as the \ion{O}{8} edge, is required due
to the strong absorption lines seen from these ions (see \S\ref{cog})
and in order to explain the spectral curvature at low energies ($\la
2$~keV) seen in Figure~\ref{continuum}. A detailed model, accounting
for all the features and processes together, is needed in order to
use these features to constrain the properties of the absorbing gas
(H. Netzer et al., in preparation).

Motivated by suggestions for the association of dust with the warm
absorber in several AGNs (e.g., Brandt, Fabian, \& Pounds 1996;
Reynolds 1997; Komossa \& Bade 1998; Lee et al. 2001) we looked
for dust features in our X-ray spectrum. We do not detect a clear
Fe~L3 edge at 707~eV (17.54~\AA) or Fe~L2 edge at 720~eV (17.22~\AA),
implying no large amount of neutral iron is in our line of sight. The
conclusion is not strong because of other predicted strong features in
this part of the spectrum. We also looked for the much clearer evidence
of a neutral oxygen edge at 538 eV. We find no evidence for such an
edge within the S/N of our data and no clear deviation from the adopted
power-law around this region (see Figure~\ref{continuum}). This result
is in agreement with the Reynolds (1997) conclusion that there is
little dust along the line of sight to the central source of NGC\,3783.

We have looked for the $1s2s2p$ (KLL) resonances of \ion{O}{6} at 22.05
\AA\ and 21.87 \AA\ which are predicted to be strong in some highly
ionized gases (Pradhan 2000). We find no evidence for absorption by
these lines in our spectrum, though their location on the wings of
the strong \ion{O}{7} forbidden and intercombination emission lines
may hinder their detection. We set an upper limit on the EW of the
\ion{O}{6} 22.05 \AA\ absorption line to be 
70\,m\AA .

\end{multicols}
\begin{deluxetable}{ccccl}
\tablecolumns{5}
\tabletypesize{\footnotesize}
\tablewidth{0pt}
\tablecaption{Line Measurements
\label{linetab}}
\tablehead{
\colhead{Observed $\lambda$} &
\colhead{FWHM} &
\colhead{EW\tablenotemark{a}} &
\colhead{Flux\tablenotemark{b}} &
\colhead{Ion name and transition rest-frame wavelength} \\
\colhead{(\AA)} &
\colhead{(\kms)} &
\colhead{(m\AA)} &
\colhead{($\times10^{-6}\phcms$)} &
\colhead{(\AA)} 
}
\startdata
$ 1.567^{+0.003}_{ -0.003}$ &$3585^{+1981}_{-1981}$ &$  6.6\pm  3.8$ &$  14.7\pm  8.5$ &\ion{Fe}{25} (1.573) \\ 
$ 1.852^{+0.002}_{ -0.002}$ &$2599^{+1393}_{-1393}$ &$  3.8\pm  1.4$ &$   7.9\pm  2.9$ &\ion{Fe}{25} (1.850) \\ 
$ 2.494^{+0.005}_{-0.010}$ &$ 930^{+1386}_{ -836}$ &$   1.8\pm  1.0$ &$   3.3\pm  1.8$ &\ion{Ca}{20} (2.514) \\ 
$ 2.535^{+0.059}_{-0.002}$ &$ 882^{+1469}_{ -450}$ &$   1.3\pm  0.7$ &$   2.4\pm  1.3$ &\ion{Ca}{20} (2.549) \\ 
$ 2.571^{+0.054}_{-0.029}$ &$1457^{+ 690}_{ -464}$ &$   1.4\pm  0.7$ &$   2.5\pm  1.2$ &\ion{Ca}{19} (2.571) \\ 
$ 2.696^{+0.022}_{-0.007}$ &$2153^{+5610}_{-1514}$ &$   1.7\pm  0.8$ &$   3.1\pm  1.5$ &\ion{Ca}{19} (2.705) \\ 
$ 2.769^{+0.001}_{-0.001}$ &$ 939^{+ 341}_{ -617}$ &$   1.6\pm  0.6$ &$   2.8\pm  1.0$ & No identification    \\ 
$ 2.905^{+0.005}_{-0.010}$ &$2793^{+2794}_{-1344}$ &$   1.8\pm  1.3$ &$   3.0\pm  2.2$ & No identification   \\ 
$ 3.020^{+0.010}_{-0.010}$ &$ 197^{+ 561}_{  -39}$ &$   1.0\pm  0.7$ &$   1.7\pm  1.2$ &\ion{Ca}{20} (3.020) \\ 
$ 3.173^{+0.005}_{-0.004}$ &$ 531^{+ 978}_{ -417}$ &$   1.9\pm  0.7$ &$   3.1\pm  1.1$ &\ion{Ca}{19} (3.177) \\ 
$ 3.391^{+0.158}_{-0.068}$ &$4090^{+2707}_{-1415}$ &$   2.6\pm  1.1$ &$   4.1\pm  1.6$ & No identification   \\ 
$ 3.492^{+0.014}_{-0.053}$ &$1560^{+1017}_{-1175}$ &$   2.4\pm  0.9$ &$   3.6\pm  1.3$ & No identification   \\ 
$ 3.734^{+0.004}_{-0.004}$ &$1972^{+ 846}_{ -584}$ &$   4.1\pm  1.1$ &$   6.1\pm  1.6$ &\ion{Ar}{18} (3.733) \\ 
$ 3.820^{+0.027}_{-0.002}$ &$ 598^{+1349}_{ -390}$ &$   1.7\pm  0.7$ &$   2.5\pm  1.0$ & No identification   \\ 
$ 3.890^{+0.005}_{-0.004}$ &$1275^{+1215}_{ -805}$ &$   1.9\pm  1.0$ &$   2.8\pm  1.5$ & No identification   \\ 
$ 3.942^{+0.004}_{-0.004}$ &$2210^{+ 829}_{ -559}$ &$   5.0\pm  1.1$ &$   7.2\pm  1.6$ &\ion{Ar}{17} (3.949) \\ 
$ 3.978^{+0.002}_{-0.002}$ &$1064^{+ 584}_{ -349}$ &$   3.7\pm  1.0$ &$   5.4\pm  1.5$ &\ion{S }{16} (3.991), \ion{S }{15} (3.998) \\ 
$ 4.076^{+0.004}_{-0.005}$ &$2199^{+ 980}_{ -599}$ &$   3.8\pm  1.1$ &$   5.4\pm  1.6$ &\ion{S }{15} (4.088) \\ 
$ 4.290^{+0.004}_{-0.003}$ &$1381^{+ 825}_{ -412}$ &$   2.7\pm  1.0$ &$   3.7\pm  1.4$ &\ion{S }{15} (4.299) \\ 
$ 4.351^{+0.003}_{-0.003}$ &$1185^{+ 487}_{ -318}$ &$   2.6\pm  0.9$ &$   3.6\pm  1.2$ & No identification   \\ 
$ 4.581^{+0.004}_{-0.003}$ &$ 577^{+ 887}_{ -428}$ &$   1.8\pm  1.1$ &$   2.4\pm  1.5$ & No identification   \\ 
$ 4.612^{+0.018}_{-0.011}$ &$3394^{+3442}_{-2307}$ &$   2.5\pm  1.2$ &$   3.2\pm  1.5$ & No identification   \\ 
$ 4.718^{+0.002}_{-0.002}$ &$1486^{+ 330}_{ -257}$ &$  10.7\pm  1.2$ &$  13.9\pm  1.6$ &\ion{S }{16} (4.729) \\ 
$ 4.932^{+0.009}_{-0.012}$ &$3364^{+2212}_{-1183}$ &$   3.9\pm  1.4$ &$   5.0\pm  1.8$ &\ion{Si}{14} (4.947) \\ 
$ 5.028^{+0.001}_{-0.001}$ &$1143^{+ 210}_{ -174}$ &$   9.2\pm  1.2$ &$  11.7\pm  1.5$ &\ion{S }{15} (5.039) \\ 
$ 5.077^{+0.003}_{-0.003}$ &$ 544^{+ 459}_{ -359}$ &$   2.7\pm  0.8$ &$   3.4\pm  1.0$ &\ion{S }{14} (5.084, 5.086) \\ 
$ 5.119^{+0.001}_{-0.001}$ &$ 695^{+ 227}_{ -181}$ &$   4.0\pm  1.1$ &$   5.0\pm  1.4$ &\ion{S }{13} (5.126) \\ 
$ 5.164^{+0.007}_{-0.005}$ &$1442^{+ 660}_{-1262}$ &$   5.2\pm  1.4$ &$   6.4\pm  1.7$ &\ion{S }{12} (5.176, 5.183) \\ 
$ 5.210^{+0.001}_{-0.002}$ &$ 631^{+ 607}_{ -467}$ &$  11.1\pm  1.6$ &$  13.7\pm  2.0$ &\ion{Si}{14} (5.217), \ion{Si}{13} (5.223), \ion{S}{11} (5.234) \\ 
$ 5.279^{+0.003}_{-0.003}$ &$ 779^{+ 670}_{ -298}$ &$   3.5\pm  1.5$ &$   4.3\pm  1.8$ &\ion{Si}{13} (5.286), \ion{S}{10} (5.276) \\ 
$ 5.402^{+0.005}_{-0.005}$ &$1414^{+ 618}_{ -469}$ &$   4.5\pm  1.6$ &$   5.5\pm  2.0$ &\ion{Si}{13} (5.405) \\ 
$ 5.671^{+0.002}_{-0.002}$ &$ 799^{+ 252}_{ -173}$ &$   6.3\pm  1.2$ &$   7.6\pm  1.4$ &\ion{Si}{13} (5.681) \\ 
$ 5.739^{+0.030}_{-0.050}$ &$ 208^{+ 365}_{ -167}$ &$   1.3\pm  0.9$ &$   1.5\pm  1.0$ &\ion{Al}{13} (5.739) \\ 
$ 6.030^{+0.003}_{-0.003}$ &$2090^{+ 459}_{ -374}$ &$   4.6\pm  0.9$ &$   4.9\pm  1.0$ &\ion{Al}{13} (6.053) \\ 
$ 6.168^{+0.001}_{-0.001}$ &$1243^{+  79}_{  -73}$ &$  20.5\pm  0.8$ &$  20.7\pm  0.8$ &\ion{Si}{14} (6.182) \\ 
$ 6.634^{+0.001}_{-0.001}$ &$ 856^{+  79}_{  -71}$ &$  14.9\pm  0.7$ &$  13.7\pm  0.6$ &\ion{Si}{13} (6.648) \\ 
$ 6.708^{+0.001}_{-0.001}$ &$ 640^{+ 160}_{ -123}$ &$   5.9\pm  0.7$ &$   5.3\pm  0.6$ &\ion{Si}{12} (6.718 ), \ion{Mg}{12} (6.738) \\ 
$ 6.770^{+0.001}_{-0.001}$ &$ 465^{+  83}_{  -78}$ &$   4.8\pm  0.6$ &$   4.3\pm  0.5$ &\ion{Si}{11} (6.778) \\ 
$ 6.844^{+0.001}_{-0.001}$ &$ 925^{+ 102}_{  -89}$ &$  10.9\pm  0.8$ &$   9.6\pm  0.7$ &\ion{Si}{10} (6.854, 6864) \\ 
$ 6.913^{+0.002}_{-0.002}$ &$1600^{+ 323}_{ -264}$ &$   7.1\pm  0.7$ &$   6.2\pm  0.6$ &\ion{Si}{9} (6.939, 6.923) \\ 
$ 6.984^{+0.001}_{-0.002}$ &$ 462^{+ 229}_{ -102}$ &$   4.3\pm  0.8$ &$   3.7\pm  0.7$ &\ion{Si}{8} (6.999) \\ 
$ 7.026^{+0.071}_{-0.015}$ &$ 540^{+ 215}_{ -215}$ &$   1.7\pm  0.6$ &$   1.5\pm  0.5$ &\ion{Si}{7} (7.063) \\ 
$ 7.089^{+0.002}_{-0.002}$ &$1182^{+ 210}_{ -166}$ &$   8.6\pm  0.9$ &$   7.3\pm  0.8$ &\ion{Mg}{12} (7.106) \\ 
$ 7.159^{+0.001}_{-0.001}$ &$ 524^{+ 129}_{ -105}$ &$   5.4\pm  0.8$ &$   4.6\pm  0.7$ &\ion{Al}{13} (7.173) \\ 
$ 7.294^{+0.002}_{-0.003}$ &$ 661^{+ 214}_{ -353}$ &$   1.6\pm  0.7$ &$   1.3\pm  0.6$ &\ion{Mg}{11} (7.310) \\ 
$ 7.459^{+0.001}_{-0.001}$ &$ 540^{+ 129}_{ -100}$ &$   7.6\pm  1.1$ &$   6.2\pm  0.9$ &\ion{Mg}{11} (7.473) \\ 
$ 7.628^{+0.005}_{-0.354}$ &$ 421^{+ 676}_{ -312}$ &$   2.5\pm  0.7$ &$   2.0\pm  0.6$ &\ion{Fe}{21} (7.635, 7.639),  \ion{Fe}{22} (7.637, 7.641, 7.645) \\ 
$ 7.673^{+0.004}_{-0.003}$ &$ 798^{+ 350}_{ -246}$ &$   3.3\pm  0.9$ &$   2.6\pm  0.7$ &\ion{Fe}{21} (7.689),  \ion{Fe}{22} (7.687) \\ 
$ 7.749^{+0.003}_{-0.003}$ &$1173^{+ 313}_{ -250}$ &$   6.3\pm  1.0$ &$   5.0\pm  0.8$ &\ion{Al}{12} (7.757) \\ 
$ 7.837^{+0.001}_{-0.001}$ &$ 679^{+ 170}_{ -122}$ &$  10.6\pm  1.0$ &$   8.2\pm  0.8$ &\ion{Mg}{11} (7.851) near the \ion{Al}{12} triplet \\ 
$ 7.908^{+0.030}_{-0.030}$ &$ 145^{+ 246}_{  -60}$ &$   4.0\pm  1.0$ &$   3.1\pm  0.8$ &\ion{Fe}{21} (7.914), \ion{Fe}{22} (7.898), \ion{Fe}{23} (7.914), \\ 
$ 7.985^{+0.005}_{-0.033}$ &$1392^{+ 755}_{ -396}$ &$   9.4\pm  1.2$ &$   7.0\pm  0.9$ &\ion{Fe}{24} (7.982) \\ 
$ 8.088^{+0.002}_{-0.001}$ &$ 291^{+ 152}_{ -185}$ &$   2.9\pm  0.9$ &$   2.1\pm  0.7$ &\ion{Fe}{22} (8.097) \\ 
$ 8.145^{+0.003}_{-0.003}$ &$ 636^{+ 227}_{ -186}$ &$   5.3\pm  1.0$ &$   3.8\pm  0.7$ &\ion{Fe}{21} (8.159) \\ 
$ 8.295^{+0.002}_{-0.002}$ &$ 556^{+ 265}_{ -213}$ &$   6.7\pm  1.2$ &$   4.6\pm  0.8$ &\ion{Fe}{23} (8.305) \\ 
$ 8.403^{+0.001}_{-0.001}$ &$ 943^{+  81}_{  -76}$ &$  25.1\pm  1.4$ &$  16.8\pm  0.9$ &\ion{Mg}{12} (8.421) \\ 
$ 8.555^{+0.002}_{-0.002}$ &$ 782^{+ 151}_{ -137}$ &$   5.6\pm  1.1$ &$   3.6\pm  0.7$ &\ion{Fe}{21} (8.580) \\ 
$ 8.709^{+0.005}_{-0.005}$ &$ 907^{+ 358}_{ -249}$ &$   8.1\pm  1.4$ &$   5.0\pm  0.9$ &\ion{Fe}{22} (8.720) \\ 
$ 8.797^{+0.005}_{-0.004}$ &$ 622^{+ 348}_{ -193}$ &$   6.3\pm  1.3$ &$   3.8\pm  0.8$ &\ion{Fe}{21} (8.842, 8.829), \ion{Fe}{20} (8.825, 8.824, 8.822) \\ 
$ 8.967^{+0.002}_{-0.002}$ &$1105^{+ 187}_{ -158}$ &$   8.8\pm  1.3$ &$   5.1\pm  0.8$ &\ion{Fe}{22} (8.982) \\ 
$ 9.059^{+0.002}_{-0.002}$ &$ 479^{+ 136}_{  -99}$ &$   3.5\pm  0.9$ &$   2.0\pm  0.5$ &\ion{Fe}{20} (9.080) \\ 
\\
\\
\\
\\
\\
\\
$ 9.152^{+0.001}_{-0.001}$ &$ 460^{+  77}_{  -66}$ &$  16.7\pm  1.0$ &$   9.4\pm  0.6$ &\ion{Mg}{11} (9.169) \\ 
$ 9.225^{+0.003}_{-0.003}$ &$ 776^{+ 303}_{ -188}$ &$   7.2\pm  1.1$ &$   4.1\pm  0.6$ &\ion{Ne}{10} (9.246) blend with \ion{Mg}{11} triplet \\ 
$ 9.267^{+0.001}_{-0.001}$ &$ 571^{+ 113}_{  -89}$ &$  10.2\pm  1.2$ &$   5.7\pm  0.7$ &\ion{Ne}{10} (9.291) blend with \ion{Mg}{11} triplet \\ 
$ 9.299^{+0.009}_{-0.003}$ &$  96^{+  99}_{  -23}$ &$  -4.1\pm  1.4$ &$   2.3\pm  0.8$ &\ion{Mg}{11} (9.300), \ion{Ne}{10} (9.291) \\ 
$ 9.341^{+0.002}_{-0.002}$ &$ 497^{+ 139}_{ -116}$ &$   7.1\pm  1.2$ &$   3.9\pm  0.7$ &\ion{Ne}{10} (9.362) deblended from \ion{Mg}{9} (9.378) \\ 
$ 9.369^{+0.002}_{-0.002}$ &$ 778^{+ 122}_{ -115}$ &$  10.2\pm  1.1$ &$   5.6\pm  0.6$ &\ion{Mg}{9} (9.378) deblended from \ion{Ne}{10} (9.362) \\ 
$ 9.466^{+0.002}_{-0.003}$ &$1200^{+ 255}_{ -205}$ &$  20.1\pm  1.5$ &$  11.0\pm  0.8$ &\ion{Ne}{10} (9.481), \ion{Fe}{21} (9.483) \\ 
$ 9.689^{+0.003}_{-0.003}$ &$1272^{+ 208}_{ -192}$ &$  30.1\pm  1.6$ &$  16.2\pm  0.9$ &\ion{Ne}{10} (9.708) \\ 
$ 9.740^{+0.003}_{-0.004}$ &$ 526^{+ 497}_{ -226}$ &$  -2.8\pm  1.4$ &$   1.5\pm  0.7$ &\ion{Ne}{10} (9.708) \\ 
$ 9.769^{+0.002}_{-0.002}$ &$ 532^{+ 119}_{ -105}$ &$   3.9\pm  1.3$ &$   2.1\pm  0.7$ &\ion{Fe}{18} (9.804) \\ 
$ 9.829^{+0.001}_{-0.001}$ &$ 567^{+ 102}_{  -90}$ &$   8.4\pm  1.5$ &$   4.5\pm  0.8$ &\ion{Fe}{19} (9.856) \\ 
$ 9.976^{+0.003}_{-0.003}$ &$1341^{+ 217}_{ -186}$ &$  19.9\pm  2.0$ &$  10.4\pm  1.0$ &\ion{Fe}{20} (9.999, 10.001, 10.006) \\ 
$10.030^{+0.003}_{-0.003}$ &$1330^{+ 240}_{ -180}$ &$  22.8\pm  1.9$ &$  11.9\pm  1.0$ &\ion{Fe}{20} (10.040, 10.042,  10.054, 10.060) \\ 
$10.100^{+0.002}_{-0.002}$ &$1146^{+ 137}_{ -120}$ &$  18.1\pm  2.0$ &$   9.4\pm  1.0$ &\ion{Fe}{17} (10.112) \\ 
$10.214^{+0.001}_{-0.001}$ &$ 893^{+  79}_{  -73}$ &$  19.6\pm  2.0$ &$  10.0\pm  1.0$ &\ion{Ne}{10} (10.239) \\ 
$10.254^{+0.008}_{-0.005}$ &$ 850^{+ 494}_{ -279}$ &$  -6.3\pm  2.3$ &$   3.2\pm  1.2$ &\ion{Ne}{10} (10.239) \\ 
$10.338^{+0.002}_{-0.002}$ &$ 953^{+ 133}_{ -121}$ &$  12.5\pm  2.2$ &$   6.3\pm  1.1$ &\ion{Fe}{18} (10.361, 10.363, 10.365) \\ 
$10.497^{+0.003}_{-0.003}$ &$1028^{+ 187}_{ -147}$ &$  16.3\pm  2.1$ &$   8.1\pm  1.0$ &\ion{Fe}{17} (10.504) \\ 
$10.623^{+0.004}_{-0.004}$ &$2091^{+ 378}_{ -304}$ &$  48.0\pm  2.4$ &$  23.5\pm  1.2$ &\ion{Fe}{19} (10.650, 10.642, 10.630, 10.631, 10.641), \ion{Fe}{17} (10.657) \\ 
$10.752^{+0.004}_{-0.003}$ &$1200^{+ 257}_{ -211}$ &$  13.7\pm  2.0$ &$   6.6\pm  1.0$ &\ion{Ne}{ 9} (10.764),  \ion{Fe}{17} (10.770) \\ 
$10.799^{+0.001}_{-0.001}$ &$ 743^{+ 127}_{  -99}$ &$  18.2\pm  1.7$ &$   8.7\pm  0.8$ &\ion{Fe}{19} (10.828) \\ 
$10.975^{+0.002}_{-0.002}$ &$1720^{+ 100}_{  -95}$ &$  45.5\pm  3.2$ &$  21.3\pm  1.5$ &\ion{Ne}{ 9} (11.000), \ion{Fe}{23} (10.981, 11.019), \ion{Fe}{17} (11.026) \\ 
$11.113^{+0.002}_{-0.002}$ &$ 652^{+ 147}_{ -120}$ &$   9.0\pm  2.2$ &$   4.1\pm  1.0$ &\ion{Fe}{17} (11.131) \\ 
$11.229^{+0.002}_{-0.002}$ &$ 696^{+ 237}_{ -190}$ &$  11.8\pm  2.5$ &$   5.3\pm  1.1$ &\ion{Fe}{17} (11.254) \\ 
$11.297^{+0.002}_{-0.002}$ &$1349^{+ 165}_{ -148}$ &$  31.0\pm  3.0$ &$  13.8\pm  1.3$ &\ion{Fe}{18} (11.326, 11.319,  11.315) \\ 
$11.399^{+0.003}_{-0.003}$ &$1742^{+ 235}_{ -206}$ &$  28.3\pm  2.7$ &$  12.4\pm  1.2$ &\ion{Fe}{22} (11.427), \ion{Fe}{18} (11.423) \\ 
$11.468^{+0.003}_{-0.002}$ &$ 695^{+ 174}_{ -132}$ &$  15.5\pm  2.4$ &$   6.7\pm  1.0$ &\ion{Fe}{22} (11.492, 11.505) \\ 
$11.513^{+0.001}_{-0.001}$ &$ 845^{+  98}_{  -87}$ &$  27.6\pm  3.0$ &$  11.8\pm  1.3$ &\ion{Ne}{ 9} (11.547) \\ 
$11.745^{+0.002}_{-0.008}$ &$1315^{+ 146}_{ -418}$ &$  37.6\pm  3.5$ &$  15.1\pm  1.4$ &\ion{Fe}{22} (11.780) deblended from \ion{Fe}{21} (11.825) \\ 
$11.805^{+0.056}_{-0.028}$ &$ 875^{+4140}_{ -254}$ &$  13.8\pm  2.9$ &$   5.4\pm  1.1$ &\ion{Fe}{21} (11.825) deblended from \ion{Fe}{22} (11.780)  \\ 
$11.934^{+0.004}_{-0.004}$ &$1929^{+ 348}_{ -271}$ &$  30.2\pm  3.7$ &$  11.8\pm  1.4$ &\ion{Fe}{21} (11.952, 11.973) \\ 
$12.010^{+0.004}_{-0.004}$ &$ 415^{+ 276}_{ -132}$ &$   4.4\pm  2.1$ &$   1.7\pm  0.8$ &\ion{Fe}{21} (12.056, 12.047) \\ 
$12.096^{+0.002}_{-0.002}$ &$1123^{+  86}_{  -78}$ &$  38.1\pm  2.9$ &$  14.7\pm  1.1$ &\ion{Ne}{10} (12.134), \ion{Fe}{17} (12.124) \\ 
$12.134^{+0.005}_{-0.004}$ &$ 363^{+ 197}_{ -292}$ &$  -3.6\pm  2.4$ &$   1.4\pm  0.9$ &\ion{Ne}{10} (12.134), \ion{Fe}{17} (12.124) \\ 
$12.169^{+0.003}_{-0.002}$ &$ 350^{+ 200}_{  -99}$ &$   5.0\pm  2.3$ &$   1.9\pm  0.9$ & \ion{Ne}{7} (12.175) \\ 
$12.243^{+0.003}_{-0.007}$ &$1229^{+ 154}_{ -320}$ &$  25.8\pm  2.4$ &$   9.8\pm  0.9$ &\ion{Fe}{17} (12.266) deblended from \ion{Fe}{21} (12.284) \\ 
$12.267^{+0.002}_{-0.002}$ &$ 346^{+ 264}_{ -124}$ &$  21.4\pm  2.6$ &$   8.1\pm  1.0$ &\ion{Fe}{21} (12.284) deblended from \ion{Fe}{17} (12.266) \\ 
$12.404^{+0.002}_{-0.002}$ &$ 512^{+ 154}_{ -112}$ &$  10.0\pm  2.9$ &$   3.7\pm  1.1$ & no identification \\ 
$12.527^{+0.019}_{-0.007}$ &$ 950^{+1020}_{ -394}$ &$  15.5\pm  2.5$ &$   5.7\pm  0.9$ &\ion{Fe}{20} (12.576) deblended from \ion{Fe}{20} (12.588) \\ 
$12.560^{+0.004}_{-0.004}$ &$ 614^{+ 203}_{ -289}$ &$  22.9\pm  2.7$ &$   8.3\pm  1.0$ &\ion{Fe}{20} (12.588) deblended from \ion{Fe}{20} (12.576) \\ 
$12.632^{+0.004}_{-0.004}$ &$ 729^{+ 283}_{ -197}$ &$  10.1\pm  3.4$ &$   3.6\pm  1.2$ & no identification \\ 
$12.727^{+0.004}_{-0.004}$ &$ 728^{+ 281}_{ -193}$ &$  11.2\pm  3.7$ &$   3.9\pm  1.3$ &\ion{Fe}{20} (12.754) \\ 
$12.804^{+0.002}_{-0.002}$ &$1389^{+ 134}_{ -114}$ &$  56.7\pm  5.6$ &$  19.5\pm  1.9$ &\ion{Fe}{20} (12.846, 12.864)  \\ 
$12.907^{+0.003}_{-0.003}$ &$1991^{+ 221}_{ -181}$ &$  66.6\pm  5.8$ &$  22.3\pm  1.9$ &\ion{Fe}{20} (12.921, 12.903), \ion{Fe}{19} (12.946, 12.933, 12.932) \\ 
$13.014^{+0.005}_{-0.005}$ &$1429^{+ 313}_{ -245}$ &$  31.1\pm  5.9$ &$  10.1\pm  1.9$ &\ion{Fe}{19} (13.022),
\ion{Fe}{21} (13.043), \ion{Fe}{20} (13.061) \\ 
$13.299^{+0.005}_{-0.004}$ &$ 513^{+ 350}_{ -182}$ &$  14.4\pm  3.8$ &$   4.2\pm  1.1$ &\ion{Fe}{18} (13.323) \\ 
$13.330^{+0.004}_{-0.006}$ &$ 465^{+ 477}_{ -171}$ &$  11.0\pm  3.5$ &$   3.2\pm  1.0$ &\ion{Fe}{18} (13.346) \\ 
$13.402^{+0.006}_{-0.005}$ &$1869^{+ 479}_{ -298}$ &$  60.2\pm  7.0$ &$  16.7\pm  1.9$ &\ion{Ne}{ 9} (13.447), \ion{Fe}{19} (13.423, 13.375), \ion{Fe}{18} (13.405) \\ 
$13.487^{+0.004}_{-0.004}$ &$1527^{+ 240}_{ -199}$ &$  55.3\pm  8.1$ &$  14.5\pm  2.1$ &\ion{Fe}{19} (13.518, 13.497) \\ 
$13.566^{+0.002}_{-0.002}$ &$ 227^{+ 132}_{ -159}$ &$ -13.9\pm  6.2$ &$   3.5\pm  1.6$ &\ion{Ne}{ 9} (13.553) \\ 
$13.690^{+0.003}_{-0.003}$ &$ 903^{+ 120}_{ -113}$ &$ -78.1\pm 12.6$ &$  17.8\pm  2.9$ &\ion{Ne}{ 9} (13.699) \\ 
$13.786^{+0.005}_{-0.005}$ &$1167^{+ 245}_{ -203}$ &$  45.9\pm  9.4$ &$   9.7\pm  2.0$ &\ion{Fe}{17} (13.825), \ion{Fe}{20} (13.767), \ion{Fe}{19} (13.795) \\ 
$14.131^{+0.009}_{-0.010}$ &$ 676^{+ 375}_{ -243}$ &$  27.4\pm  6.2$ &$   6.7\pm  1.5$ &\ion{Fe}{18} (14.143, 14.158) \\ 
$14.175^{+0.006}_{-0.008}$ &$ 858^{+ 520}_{ -334}$ &$  24.8\pm  5.5$ &$   6.5\pm  1.4$ &\ion{Fe}{18} (14.208) \\ 
$14.232^{+0.004}_{-0.004}$ &$ 808^{+ 326}_{ -217}$ &$  20.8\pm  5.6$ &$   5.9\pm  1.6$ &\ion{Fe}{18} (14.256) \\ 
$14.346^{+0.003}_{-0.003}$ &$ 601^{+ 204}_{ -141}$ &$  16.7\pm  4.0$ &$   5.6\pm  1.3$ &\ion{Fe}{18} (14.373) \\ 
$14.435^{+0.005}_{-0.004}$ &$ 833^{+ 283}_{ -173}$ &$  18.9\pm  4.2$ &$   7.0\pm  1.6$ &\ion{O }{ 8} (14.454) \\ 
$14.460^{+0.008}_{-0.008}$ &$1096^{+6516}_{ -367}$ &$  14.7\pm  3.8$ &$   5.7\pm  1.5$ &\ion{Fe}{18} (14.483) \\ 
$14.510^{+0.002}_{-0.002}$ &$ 832^{+ 138}_{ -107}$ &$  34.5\pm  5.6$ &$  14.1\pm  2.3$ &\ion{ O}{ 8} (14.524), \ion{Fe}{18} (14.534) \\ 
$14.561^{+0.006}_{-0.008}$ &$ 336^{+1537}_{ -231}$ &$   8.5\pm  3.7$ &$   3.7\pm  1.6$ &\ion{Fe}{18} (14.571) \\ 
$14.615^{+0.003}_{-0.003}$ &$ 965^{+ 182}_{ -144}$ &$  32.4\pm  5.5$ &$  14.0\pm  2.4$ &\ion{O }{ 8} (14.634), \ion{Fe}{18} (14.616) \\ 
$14.800^{+0.004}_{-0.004}$ &$1237^{+ 210}_{ -173}$ &$  45.8\pm  8.1$ &$  18.8\pm  3.3$ &\ion{O }{ 8} (14.821) \\ 
\\
\\
\\
\\
\\
$14.942^{+0.002}_{-0.003}$ &$ 367^{+ 202}_{ -255}$ &$  15.0\pm  4.4$ &$   6.0\pm  1.8$ &\ion{Fe}{19} (14.961) \\ 
$14.980^{+0.003}_{-0.003}$ &$ 593^{+ 134}_{ -105}$ &$  26.2\pm  5.3$ &$  10.3\pm  2.1$ &\ion{Fe}{17} (15.014) \\ 
$15.153^{+0.006}_{-0.007}$ &$1489^{+ 414}_{ -308}$ &$  38.5\pm  7.7$ &$  14.3\pm  2.9$ &\ion{O }{ 8} (15.176) \\ 
$15.231^{+0.001}_{-0.002}$ &$ 259^{+ 143}_{ -119}$ &$  14.9\pm  4.3$ &$   5.4\pm  1.6$ &\ion{Fe}{17} (15.261) \\ 
$15.970^{+0.004}_{-0.005}$ &$ 764^{+ 235}_{ -162}$ &$  36.7\pm  9.2$ &$   9.5\pm  2.4$ &\ion{O }{ 8} (16.006) \\ 
$16.017^{+0.008}_{-0.007}$ &$ 494^{+ 304}_{ -185}$ &$ -11.7\pm  8.7$ &$   2.9\pm  2.1$ &\ion{O }{ 8} (16.006) \\ 
$16.695^{+0.008}_{-0.008}$ &$ 956^{+ 307}_{ -232}$ &$ -26.0\pm 18.7$ &$   5.1\pm  3.6$ & Radiative recombination continuum at the edge of \ion{O }{ 7} (16.771) \\ 
$17.161^{+0.004}_{-0.005}$ &$ 738^{+ 256}_{ -192}$ &$  28.7\pm  8.2$ &$  13.0\pm  3.7$ &\ion{O }{ 7} (17.200) \\ 
$17.351^{+0.004}_{-0.005}$ &$ 962^{+ 204}_{ -156}$ &$  40.7\pm  8.1$ &$  18.6\pm  3.7$ &\ion{O }{ 7} (17.396) \\ 
$17.719^{+0.005}_{-0.005}$ &$ 853^{+ 199}_{ -171}$ &$  37.3\pm  8.5$ &$  16.7\pm  3.8$ &\ion{O }{ 7} (17.768) \\ 
$17.778^{+0.007}_{-0.001}$ &$ 133^{+  49}_{  -39}$ &$ -10.9\pm  9.6$ &$   4.9\pm  4.3$ &\ion{O }{ 7} (17.768) \\ 
$18.572^{+0.012}_{-0.012}$ &$ 889^{+ 373}_{ -477}$ &$  31.0\pm 10.9$ &$  13.4\pm  4.7$ &\ion{O }{ 7} (18.627) \\ 
$18.630^{+0.023}_{-0.008}$ &$ 118^{+ 117}_{  -72}$ &$ -20.6\pm 11.9$ &$   8.9\pm  5.1$ &\ion{O }{ 7} (18.627) \\ 
$18.906^{+0.007}_{-0.007}$ &$ 866^{+ 199}_{ -164}$ &$  53.6\pm 15.9$ &$  22.8\pm  6.8$ &\ion{O }{ 8} (18.969) \\ 
$18.968^{+0.004}_{-0.004}$ &$ 462^{+ 110}_{ -108}$ &$ -54.7\pm 17.4$ &$  23.2\pm  7.4$ &\ion{O }{ 8} (18.969) \\ 
$19.297^{+0.016}_{-0.016}$ &$1040^{+ 582}_{ -353}$ &$  30.1\pm 17.4$ &$  12.6\pm  7.3$ &\ion{N }{ 7} (19.361) \\ 
$19.793^{+0.006}_{-0.006}$ &$ 284^{+ 185}_{ -207}$ &$  13.4\pm  9.2$ &$   5.5\pm  3.8$ &\ion{N }{ 7} (19.826) \\ 
$20.871^{+0.012}_{-0.014}$ &$ 572^{+ 427}_{ -349}$ &$  33.9\pm 24.7$ &$  13.3\pm  9.7$ &\ion{N }{ 7} (20.910) \\ 
$20.922^{+0.020}_{-0.011}$ &$ 352^{+ 718}_{ -269}$ &$ -33.0\pm 28.0$ &$  12.9\pm 10.9$ &\ion{N }{ 7} (20.910) \\ 
$21.551^{+0.008}_{-0.016}$ &$ 365^{+ 383}_{ -149}$ &$  40.1\pm 34.6$ &$  15.4\pm 13.3$ &\ion{O }{ 7} (21.602) \\ 
$21.605^{+0.007}_{-0.006}$ &$ 637^{+ 210}_{ -165}$ &$-120.1\pm 62.3$ &$  45.8\pm 23.8$ &\ion{O }{ 7} (21.602) \\ 
$21.834^{+0.007}_{-0.008}$ &$ 558^{+ 350}_{ -177}$ &$ -93.1\pm 48.2$ &$  35.2\pm 18.2$ &\ion{O }{ 7} (21.807) \\ 
$22.089^{+0.006}_{-0.008}$ &$ 927^{+ 361}_{ -370}$ &$-270.4\pm 73.2$ &$ 101.3\pm 27.4$ &\ion{O }{ 7} (22.101) \\ 
$22.376^{+0.005}_{-0.008}$ &$ 208^{+ 354}_{ -205}$ &$ -39.5\pm 39.8$ &$  10.8\pm 10.9$ & Radiative recombination continuum at the edge of \ion{N}{6} (22.458)\\ 
$22.856^{+0.231}_{-0.049}$ &$ 938^{+1084}_{ -812}$ &$ -90.9\pm 66.9$ &$  33.2\pm 24.4$ & Instrument feature \\ 
$23.056^{+0.009}_{-0.008}$ &$ 470^{+ 416}_{ -258}$ &$ -80.6\pm 44.4$ &$  29.2\pm 16.1$ & Instrument feature \\ 
$23.290^{+0.014}_{-0.054}$ &$ 338^{+2650}_{  -59}$ &$  47.0\pm 35.6$ &$  16.9\pm 12.8$ &\ion{N }{ 6} (23.277) \\ 
$24.708^{+0.015}_{-0.018}$ &$ 951^{+ 627}_{ -330}$ &$  72.6\pm 34.9$ &$  24.9\pm 12.0$ &\ion{N }{ 7} (24.781) \\ 
$24.768^{+0.216}_{-0.009}$ &$  83^{+  57}_{  -59}$ &$ -16.3\pm 35.7$ &$   5.6\pm 12.3$ &\ion{N }{ 7} (24.781) \\ 
$24.840^{+0.012}_{-0.013}$ &$ 331^{+ 383}_{ -258}$ &$  21.6\pm 19.7$ &$   7.4\pm  6.7$ &\ion{N }{ 6} (24.898) \\ 
$24.896^{+0.011}_{-0.011}$ &$  95^{+ 120}_{  -76}$ &$ -18.5\pm 25.4$ &$   6.3\pm  8.6$ &\ion{N }{ 6} (24.898) \\ 
$25.974^{+0.015}_{-0.239}$ &$ 369^{+ 669}_{ -300}$ &$  32.5\pm 36.3$ &$  10.7\pm 12.0$ &\ion{C }{ 6} (26.026) \\ 
$26.028^{+0.260}_{-0.260}$ &$ 322^{+ 469}_{ -322}$ &$ -24.4\pm 34.3$ &$   8.0\pm 11.2$ &\ion{C }{ 6} (26.026) \\ 
\enddata
\tablenotetext{a}{Negative signs before the EW indicate an emission line.}
\tablenotetext{b}{Absorbed flux for the absorption lines and emitted flux
for the emission lines.}
\end{deluxetable}
\begin{multicols}{2}

\noindent Assuming a single
absorbing system with a Doppler velocity of 350 \kms\ (see \S\ref{cog}
below) an upper limit for the column density of \ion{O}{6} is $10^{17}$
\cmii . This upper limit is in agreement with the simultaneous {\it
FUSE} observations of the \ion{O}{6} absorption lines (J. R. Gabel
et al., in preparation) and with the models presented by Kraemer et
al. (2001) and Kaspi et al. (2001).

Kaspi et al. (2001 \S6.2) argued against the possibility that the
absorption features they detected arise from intervening material in our
Galaxy (in particular the high-velocity cloud seen toward NGC\,3783;
Lu et al. 1998). The same considerations still hold, and the current
paper confirms that the absorption is indeed intrinsic to NGC\,3783.
The high S/N of our spectrum allows us to look for possible Galactic or
intergalactic absorption at zero redshift. The most probable line is
\ion{O}{7} (21.602~\AA) which, in Figure~\ref{xrayspec}, would be at
21.393 \AA. A feature is seen at this wavelength, but it is too weak
(${\rm EW}=14\pm16$~m\AA) to ascribe any significance to it.
Further detailed discussion and modeling of the multiwavelength Galactic
absorption toward NGC\,3783 will be discussed by S. Mathur et al. (in
preparation).

\subsection{Correlations Between Line Properties}
\label{correl}

Given that we observe absorption lines spanning a large range of energy
and ionization potential, it is interesting to look for correlations
between the line properties. However, as mentioned above most of
the lines in Table~\ref{linetab} are blended, some have low S/N,
and many are unresolved by the HETGS. Out of the 135 absorption
features listed in Table~\ref{linetab}, we chose the 42 lines which
are not blended and are strong enough for accurate velocity-shift
and FWHM measurements (EW$\ga 10$~m\AA). We first deconvolved the
instrumental FWHMs from the measured FWHMs to obtain true FWHM values,
i.e., ${\rm FWHM}^2_{\rm true} = {\rm FWHM}^2_{\rm measured} - {\rm
FWHM}^2_{\rm instrument}$. We find that 31 lines are resolved, i.e.,
${\rm FWHM}_{\rm measured} > {\rm FWHM}_{\rm instrument}$, and that 18
are resolved at the 3$\sigma$ level. We list the 42 lines with their
velocity-shift and FWHM$_{\rm true}$ in Table~3. The
wavelength accuracy of the inner-shell lines is not as good as
the H-like and He-like lines; therefore, they were not included in
this table.  We checked if the blueshifts and FWHMs are correlated
with wavelength. We find no such correlations. This confirms that
our blueshift and FWHM  
%
\centerline{\includegraphics[width=8.5cm]{f6.eps}}
\figcaption{Broad absorption feature from blended inner-shell 2p-3d
absorption lines of Fe M-shell ions forming the UTA. Data are the same as
shown in Figure~\protect\ref{xrayspec} focusing on the 15--18~\AA\ range.
\label{uta} }
\centerline{}
\centerline{}
%
\noindent measurements are not affected by systematic
instrumental effects.  We also do not find any correlation between
the blueshift and FWHM of the different lines.

Several studies have found the widths of the optical narrow emission
lines in NGC\,3783 and their blueshifts relative to the systemic
velocity to correlate with the ionization potentials needed to
create the relevant ions (e.g., Pelat, Alloin, \& Fosbury 1981;
Evans 1988). This has motivated us to search for similar trends
in the many absorption lines seen in the X-ray as they span a
large range of ionization potential (obviously, we do not expect
a simple dynamical connection between the optical narrow emission
lines and the X-ray absorption lines).  Figure~\ref{ipot_rel} shows
the blueshift and FWHM$_{\rm true}$ values of individual lines as a
function of the ionization potential required to produce the relevant
ions. No correlation is found between these properties. However,
an interesting trend is revealed by color-coding the different
transitions in each ion presented in Figure~\ref{ipot_rel}a. For a
given ion, the blueshifts seem to diminish from the 1$\rightarrow$2
transition to the 1$\rightarrow$3 transition and so on. Such an effect
can be caused by the filling of the absorption troughs with emission
which weakens as one advances along the transition series. This will
cause the 1$\rightarrow$2 transition's absorption to be filled more
on its red side than the 1$\rightarrow$3 transition's absorption,
causing the apparent wavelength shift to be larger.  Such a
phenomenon might also be related to the kinematics of outflows with
scattering. However, a full detailed model of the outflow (e.g.,
H. Netzer et al. in preparation) is needed in order to further
discuss this.  Figure~\ref{shift_vs_n} shows the correlation between
absorption-line blueshift and the transition number for \ion{O}{7}
and \ion{O}{8}. Attempting to quantify the correlation we find that the
Spearman rank-order correlation is not suitable in this case since we
have a totally concordant ranking sequence. We used the nonparametric
Kendall $\tau$ test (Press et al. 1992 \S14.6) which yields correlation
coefficients of $\tau = 1.0$ with significance levels of 0.0415 for
both ions.  We note that no correlation is found between the FWHM
and the transition number for a given ion (Figure~\ref{ipot_rel}b).

In order to look for real correlations between the blueshift and
FWHM$_{\rm true}$ values of individual lines and the ionization po-
\footnotesize 
\begin{center}
{\sc TABLE 3 \\ Properties of Unblended Strong Lines }
\vskip 4pt
\begin{tabular}{ccccl}
\hline
\hline
{} &
{} &
{Ion name and} \\
{Velocity Shift} &
{FWHM$_{\rm true}$} &
{transition rest-frame wavelength} \\
{(\kms)} &
{(\kms)} &
{(\AA)}  \\
\hline
$-699^{+ 112}_{-131}$ &$1276^{+ 373}_{ -313}$ &\ion{S }{16} (4.729) \\ 
$-636^{+  71}_{ -71}$ &$ 892^{+ 257}_{ -239}$ &\ion{S }{15} (5.039) \\ 
$-502^{+  83}_{ -84}$ &$< 840$ &\ion{Si}{13} (5.681) \\ 
$-669^{+  29}_{ -29}$ &$1099^{+  89}_{  -84}$ &\ion{Si}{14} (6.182) \\ 
$-644^{+  27}_{ -27}$ &$ 663^{+ 100}_{  -95}$ &\ion{Si}{13} (6.648) \\ 
$-714^{+  69}_{ -79}$ &$1068^{+ 230}_{ -188}$ &\ion{Mg}{12} (7.106) \\ 
$-563^{+  46}_{ -51}$ &$< 419$ &\ion{Al}{13} (7.173) \\ 
$-549^{+  45}_{ -47}$ &$< 465$ &\ion{Mg}{11} (7.473) \\ 
$-326^{+ 128}_{-124}$ &$1078^{+ 335}_{ -280}$ &\ion{Al}{12} (7.757) \\ 
$-510^{+ 105}_{-117}$ &$ 458^{+ 285}_{ -370}$ &\ion{Fe}{21} (8.159) \\ 
$-365^{+  89}_{ -75}$ &$< 697$ &\ion{Fe}{23} (8.305) \\ 
$-638^{+  31}_{ -32}$ &$ 841^{+  91}_{  -87}$ &\ion{Mg}{12} (8.421) \\ 
$-371^{+ 158}_{-163}$ &$ 808^{+ 389}_{ -295}$ &\ion{Fe}{22} (8.720) \\ 
$-498^{+  67}_{ -66}$ &$ 795^{+ 246}_{ -241}$ &\ion{Fe}{22} (8.982) \\ 
$-538^{+  37}_{ -37}$ &$ 240^{+ 127}_{ -220}$ &\ion{Mg}{11} (9.169) \\ 
$-576^{+  94}_{ -90}$ &$1217^{+ 216}_{ -203}$ &\ion{Ne}{10} (9.708) \\ 
$-362^{+  52}_{ -54}$ &$ 921^{+ 166}_{ -156}$ &\ion{Fe}{17} (10.112) \\ 
$-734^{+  33}_{ -34}$ &$ 586^{+ 115}_{ -120}$ &\ion{Ne}{10} (10.239) \\ 
$-800^{+  38}_{ -40}$ &$ 381^{+ 212}_{ -295}$ &\ion{Fe}{19} (10.828) \\ 
$-870^{+  32}_{ -32}$ &$ 597^{+ 133}_{ -132}$ &\ion{Ne}{ 9} (11.547) \\ 
$-878^{+  56}_{-196}$ &$1178^{+ 162}_{ -499}$ &\ion{Fe}{22} (11.780) \\ 
$-497^{+1421}_{-722}$ &$ 653^{+4330}_{ -442}$ &\ion{Fe}{21} (11.825) \\ 
$-565^{+  75}_{-162}$ &$1093^{+ 171}_{ -379}$ &\ion{Fe}{17} (12.266) \\ 
$-412^{+  48}_{ -48}$ &$< 241$ &\ion{Fe}{21} (12.284) \\ 
$-674^{+  83}_{ -86}$ &$< 606$ &\ion{Fe}{20} (12.588) \\ 
$-628^{+  87}_{ -93}$ &$< 853$ &\ion{Fe}{20} (12.754) \\ 
$-551^{+ 105}_{ -97}$ &$< 691$ &\ion{Fe}{18} (13.323) \\ 
$-371^{+  90}_{-136}$ &$< 788$ &\ion{Fe}{18} (13.346) \\ 
$-703^{+ 117}_{-165}$ &$ 707^{+ 583}_{ -514}$ &\ion{Fe}{18} (14.208) \\ 
$-496^{+  84}_{ -92}$ &$ 647^{+ 379}_{ -310}$ &\ion{Fe}{18} (14.256) \\ 
$-569^{+  58}_{ -63}$ &$< 647$ &\ion{Fe}{18} (14.373) \\ 
$-497^{+  47}_{ -47}$ &$ 684^{+ 163}_{ -136}$ &\ion{Fe}{18} (14.534) \\ 
$-423^{+  87}_{ -87}$ &$1147^{+ 224}_{ -190}$ &\ion{O }{ 8} (14.821) \\ 
$-377^{+  50}_{ -57}$ &$< 335$ &\ion{Fe}{19} (14.961) \\ 
$-687^{+  50}_{ -54}$ &$ 375^{+ 190}_{ -211}$ &\ion{Fe}{17} (15.014) \\ 
$-466^{+ 125}_{-131}$ &$1418^{+ 430}_{ -329}$ &\ion{O }{ 8} (15.176) \\ 
$-665^{+  78}_{ -98}$ &$ 631^{+ 272}_{ -211}$ &\ion{O }{ 8} (16.006) \\ 
$-675^{+  78}_{ -89}$ &$ 619^{+ 290}_{ -250}$ &\ion{O }{ 7} (17.200) \\ 
$-770^{+  77}_{ -81}$ &$ 877^{+ 220}_{ -175}$ &\ion{O }{ 7} (17.396) \\ 
$-827^{+  85}_{ -92}$ &$ 760^{+ 219}_{ -200}$ &\ion{O }{ 7} (17.768) \\ 
$-888^{+ 200}_{-187}$ &$ 809^{+ 399}_{ -630}$ &\ion{O }{ 7} (18.627) \\ 
$-983^{+ 105}_{-114}$ &$ 786^{+ 215}_{ -186}$ &\ion{O }{ 8} (18.969) \\ 
\hline
\end{tabular}
\end{center}
\setcounter{table}{3}
\normalsize
%
\noindent    tential,
the effect of filling of the absorption trough with emission needs to
be taken into account. Inspection of Figure~\ref{ipot_rel} suggests
that the only absorption line which looks obviously contaminated by the
emission is \ion{O}{8}\,(18.967); hence, we looked for correlations
after ignoring this line. Again no correlations were found; Spearman
rank-order correlations yield $r=0.16$ (significance level of 0.31)
for Figure~\ref{ipot_rel}a (blueshifts) and $r=0.23$ (significance
level of 0.21) for Figure~\ref{ipot_rel}b (FWHMs). The $\chi^2$ per
degree-of-freedom (dof) for a constant fit in Figure~\ref{ipot_rel}a
(blueshifts) is $\chi^2_\nu =$6.4, and in Figure~\ref{ipot_rel}b (FWHMs)
it is $\chi^2_\nu =$2.4. These results suggest that the blueshift and
the FWHM are not consistent for all the species. The mean blueshift
is $-590\pm150$ \kms, and the mean FWHM is $820\pm280$ \kms\ (the
uncertainties are the standard deviation of the measurements). We note
that correlations between velocity or FWHM and the ionization potential
were also not found in the UV band (D. M. Crenshaw et al., in preparation;
J. R. Gabel et al., in preparation).

\centerline{\includegraphics[width=8.5cm]{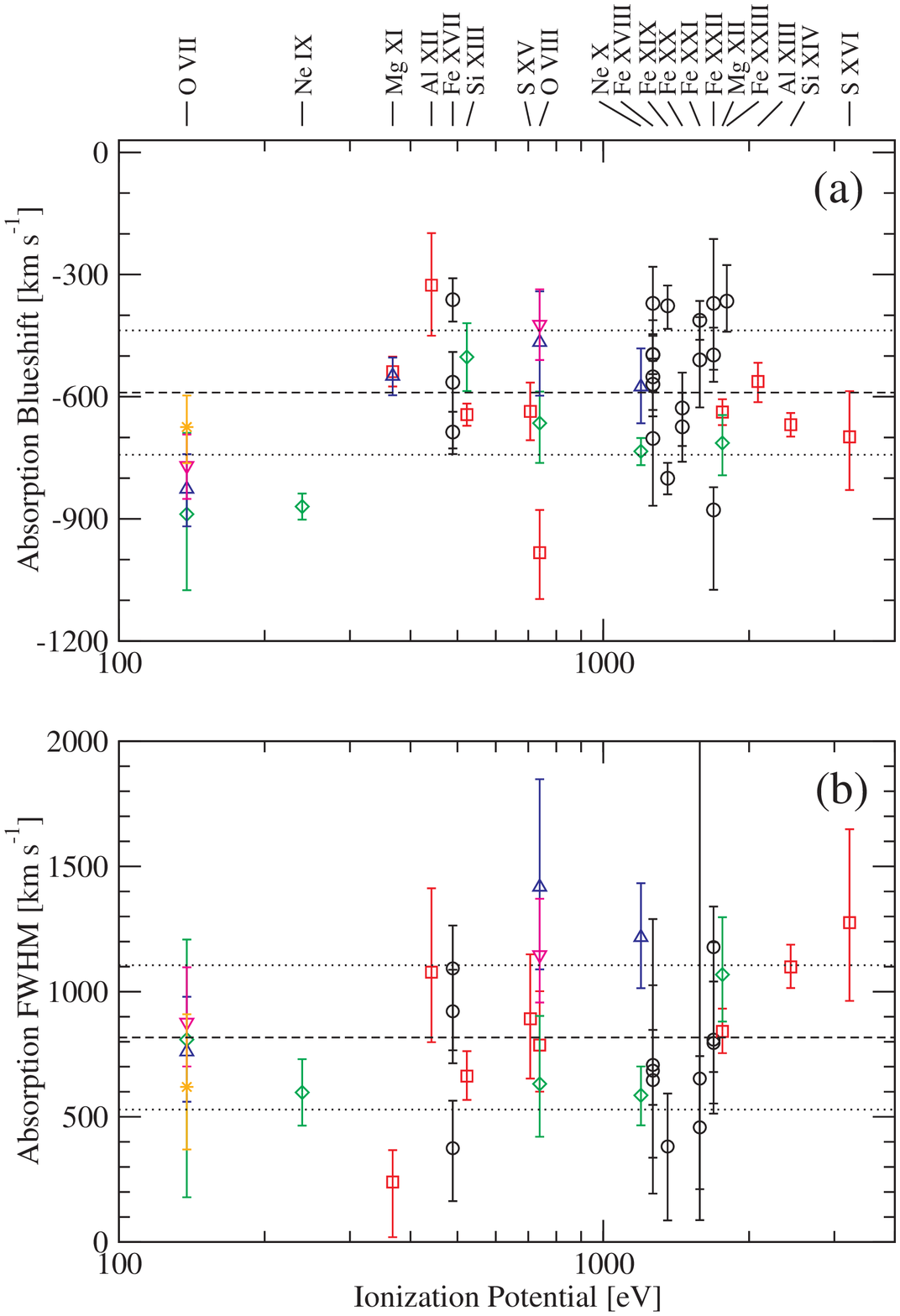}}
\figcaption{Absorption-line (a) blueshifts and (b) FWHMs as a function
of the ionization potential of the relevant ion (the energy needed to
create the ion). No trend or correlation between blueshifts or FWHMs
with the ionization potential is detected. The ions are listed above
the figure. The different transitions in the H-like and He-like ions
from the ground level to upper levels are color coded as following:
level 2 --- red squares, level 3 --- green diamonds, level 4 --- blue
triangles up, level --- 5 magenta triangles down, and level 6 ---
orange stars. A trend for the lower-level transitions to have lower
velocity blueshifts (i.e., smaller shifts from the expected wavelength
of the line) is seen. We attribute this to filling of the absorption
trough with emission as explained in the text. Average values for
each panel are plotted as horizontal dashed lines, and the standard
deviations are plotted as horizontal dotted lines.
\label{ipot_rel} }
\centerline{} 
\centerline{} 
\centerline{} 

\centerline{\includegraphics[width=8.5cm]{f8.eps}}
\figcaption{Absorption-line blueshifts as a function of $N$, where $N$
is the number for the 1$\rightarrow${$N$} transition. Squares denote
\protect\ion{O}{7} lines, and circles denote \protect\ion{O}{8} lines.
The correlation detected suggests the blueshifts diminish from the
1$\rightarrow$2 transition to the 1$\rightarrow$3 transition and so on.
\label{shift_vs_n} }

\centerline{\includegraphics[width=8.5cm]{f9.eps}}
\figcaption{Simulation of photon scattering into the absorption trough,
preventing it from becoming truly black. The solid line is the input spectrum
which has a typical shape of lines at around 20 \AA\ in our 900 ks
spectrum. Photons are drawn from this input spectrum and are scattered in
a Gaussian distribution (which has the MEG's FWHM: 0.023 \AA) along the
velocity axis. Even though the initial spectrum (solid line) is black,
the absorption in the resulting spectrum (black dots with error bars)
is smeared and thus does not appear black.
\label{bleed} }

\subsection{Curve of Growth Analysis}
\label{cog}

Curve of growth analysis (as it is formulated in Spitzer 1978) was carried
out to obtain ionic column densities using the measured absorption-line
EWs. This technique is most useful when the lines are resolved and
unsaturated. In the following we address these issues for H-like and
He-like ions in the X-ray spectrum of NGC\,3783.

\ion{O}{7} and \ion{O}{8} lines are identified up to the 1$\rightarrow$6
and 1$\rightarrow$8 transitions, respectively. Identification of
lines of such high transitions means that the first lines in each
series are probably saturated as they have much larger ($\sim 20$
times) oscillator strengths. We note that though the first lines in
each series are probably saturated, we see the effect of ``non-black''
saturation. This is probably because the resolution of the instrument
is comparable to the true line widths. In this case the convolution of
saturated lines with a line-spread function of similar width will cause
photons to scatter from the continuum (and any line emission redward
of the trough) into the absorption trough preventing it from becoming
black, even though the line is saturated. We have carried out Monte-Carlo
simulations which have confirmed the importance of this effect in our
spectra (Figure~\ref{bleed}). We have also searched for a correlation
between the EW and FWHM$_{\rm true}$ of lines for certain ions; such
a correlation is expected if the lines are saturated. Unfortunately,
due to the large uncertainties on FWHM$_{\rm true}$, we were unable
to detect such a correlation. A related problem is the filling of the
absorption trough by emission discussed in the previous section.  This is
more significant in the first lines in the series where the emission is
stronger. Also, the scattered X-ray continuum, which we constrained to
$\la 15$\% of the continuum (\S\ref{basic_line}), can contribute to the
non-black saturation we observe.

All the above suggest that reliable curve of growth analysis can
be carried out only for the lines representing transitions to high
levels which are not saturated and where line emission is weak. We
used the 1$\rightarrow$5 and 1$\rightarrow$6 transitions in both
\ion{O}{7} and \ion{O}{8}. From the FWHM$_{\rm true}$ of the lines we
estimate the Doppler velocity to be about 350~\kms . Curve of growth
analysis assuming a single absorption system yields column densities of
$N_{\rm O\,VII} = (1.1^{+0.6}_{-0.4})\times10^{18}$~\cmii\ and 
\centerline{}
\footnotesize 
\begin{center}
{\sc TABLE 4 \\  Column Density Measurements}
\vskip 4pt
\begin{tabular}{lc|lc}
\hline
\hline
{Ion} &
{$\log$\,Column [cm$^{-2}$]}  &
{Ion} &
{$\log$\,Column [cm$^{-2}$]}\\
\hline
\ion{O}{7}   & $18.04\pm0.2\phn $ & \ion{O}{8}   & $18.63\pm0.25$ \\
\ion{Mg}{11} & $17.00\pm0.4\phn $ & \ion{Mg}{12} & $17.55\pm0.1\phn $ \\
\ion{Al}{12} & $16.35\pm0.1\phn $ & \ion{Al}{13} & $16.95\pm0.4\phn $ \\
\ion{Si}{13} & $17.25\pm0.2\phn $ & \ion{Si}{14} & $17.96\pm0.2\phn $ \\
\ion{S}{15}  & $17.10\pm0.2\phn $ & \ion{S}{16}  & $17.70\pm0.2\phn $ \\
\ion{Si}{11} & $16.32\pm0.06$ & \ion{Si}{10} & $16.94\pm0.06$ \\
\ion{Si}{9} & $16.76\pm0.06$ & \ion{Si}{8} & $16.90\pm0.1\phn $ \\
\ion{Si}{7} & $16.47\pm0.1\phn $ & \ion{S}{14} & $16.32\pm0.15$ \\
\ion{S}{13}  & $16.48\pm0.15$ & \ion{S}{12}  & $16.75\pm0.15$ \\
\hline
\end{tabular}
\end{center}
\setcounter{table}{4}
\normalsize
\centerline{}
\centerline{}
\noindent 
$N_{\rm
O\,VIII} \ = \ (4.3^{+3.3}_{-1.9})\times10^{18}$ \cmii .  The corresponding
absorption-edge optical
depths are\,$\tau_{\rm O\,VII}\,(739~{\rm eV})\,= 0.26^{+0.15}_{-0.10}$\,and\,$\tau_{\rm O\,VIII}\,(871\,{\rm eV})\,=$
$0.42^{+0.33}_{-0.19}$. These values should be considered as lower
limits due to the fact that multiple systems might exist in these
ions as explained in \S\ref{two_sys}. Furthermore, these values are
inconsistent with the absorption-edge optical depths seen in the spectrum
(e.g., Figure~\ref{continuum}) which are about a factor two larger.
This can be remedied by assuming multiple absorption systems. For example,
assuming two absorption systems, each with a Doppler velocity of 120~\kms\
(comparable to the ones observed in the UV), which contribute equally to
the EW results in $\tau_{\rm O\,VII} (739~{\rm eV}) = 0.4^{+0.4}_{-0.2}$
and $\tau_{\rm O\,VIII} (871~{\rm eV}) = 1.3^{+2.9}_{-0.8}$. Further
analysis of this issue is postponed for the detailed modeling paper
(H. Netzer et al. in preparation).

The \ion{Ne}{9} and \ion{Ne}{10} absorption lines suffer the same
problems as the above-mentioned oxygen lines. An additional problem with
the neon absorption lines is that most of them are blended with iron
lines which hinders curve of growth analysis. Higher ionization ions
(H-like and He-like ions of Mg, Al, Si, and S) are probably free of
saturation and filling of the trough as we detect only the first few
lines in their series, and their emission is hardly detected. However,
if multiple unresolved systems are present in these ions the yielded
column densities would be in error. Bearing in mind these limitations we
carried out curve of growth analysis for these ions under the assumption
of one absorption system with a Doppler velocity of 350~\kms. For each
ion we used lines that are not blended (typically between 1 to 3 lines per
ion). The measured column densities are presented in Table~4.

\subsection{Multiple Absorption Components}
\label{two_sys}

Recent high-resolution X-ray studies have shown that several X-ray
absorption systems can be present in the outflows from AGNs. Two
X-ray absorption systems were reported by Sako et al. (2001)
for IRAS\,13349+2438 and by Collinge et al. (2001) for NGC\,4051.
In the X-ray spectrum of NGC\,3783 (Figures~\ref{velspec_meg} and
\ref{velspec_heg}), most absorption lines show asymmetric profiles in
that the blue wing extends to higher velocities (relative to the line
centroid) than the red wing. This could be caused by another absorption
system at higher outflow velocity and/or because the emission lines
(redward of the absorption) are filling the absorption trough. The
latter possibility, mentioned in \S\ref{correl}, was demonstrated
in the model presented by Kaspi et al. (2001) and will be discussed
by H. Netzer et al. (in preparation). The 900 ks exposure allows
us to study the possibility of multiple absorption systems. To
explore this we checked individually (and in various combinations)
all the absorption lines from the H-like and He-like ions. We show
the best cases (the 
%
\centerline{\includegraphics[width=8.5cm]{f10.eps}}
\figcaption{Velocity spectra, binned at 100 \kms, of (a) the combined
four strongest lines of \protect\ion{O}{7} (21.602, 18.627, 17.768,
and 17.396 \AA) from the MEG spectrum and (b) the combined two
strongest (un-blended with iron) lines of \protect\ion{Ne}{10}
(Ly$\beta$ 10.239 \AA\ and Ly$\gamma$ 9.708 \AA) from the HEG
spectrum. A three-Gaussian fit for each spectrum individually is
overplotted as a solid line on the data points (models 2 and 4
of Table~\protect\ref{gaussfit} for \ion{O}{7} and \ion{Ne}{10},
respectively). A three-Gaussian fit in which the positions and
the FWHMs are fixed for the two spectra together is plotted as a
dashed line (model 5 of Table~\protect\ref{gaussfit}; see text
for details). Two absorption systems are clearly detected in
\protect\ion{O}{7} and probably exist in \protect\ion{Ne}{10} as
well. Black squares with horizontal error bars indicate the velocity
shifts and FWHMs of the UV absorption systems (the UV absorbers
velocity shifts are also marked with dotted vertical lines to guide
the eye). Their agreement with the \protect\ion{O}{7} absorption and
disagreement with the \protect\ion{Ne}{10} absorption suggest that
higher ionization ions reside in a somewhat dynamically distinct
region from lower ionization ions.
\label{linesmodel} }
\centerline{}
\centerline{}
\noindent combination of best resolution and number of
counts) in Figure~\ref{linesmodel}.  The discussion below assumes
the simple approach that the data can be represented by discrete
Gaussian fits. The alternative approach of having a single medium
with gradients of ionization, density, and outflow velocity instead
of separate absorbers (e.g., Krolik \& Kriss 2001) requires detailed
modeling and will not be discussed here.

In Figure~\ref{linesmodel}a we show the velocity spectrum of the combined
four strongest lines of \ion{O}{7}, and in Figure~\ref{linesmodel}b
we show the velocity spectrum of the combined two strongest (unblended
with iron) lines of \ion{Ne}{10}. We fitted both velocity spectra with a
combination of a constant continuum and three Gaussians; two in absorption
and one in emission. The fit parameters are given in Table~\ref{gaussfit}
(model 2 for \ion{O}{7} and model 4 for \ion{Ne}{10}). For comparison,
we also list the fits with one Gaussian in absorption and one in
emission (model 1 for \ion{O}{7} and model 3 for \ion{Ne}{10}). To
check for a statistically significant improvement in fit quality when
using two absorbing Gaussians instead of only one, we used \ the \ $F$-test \
(Bevington \ \& \ Robinson \ 1992). \ \, For \ \ion{O}{7}

\end{multicols}
\begin{deluxetable}{lccccc}
\tablecolumns{6}
\tabletypesize{\footnotesize}
\tablewidth{0pt}
\tablecaption{ Gaussian Fits for Velocity-Resolved Systems
\label{gaussfit}}
\tablehead{
\colhead{Parameter} &
\colhead{Model 1} &
\colhead{Model 2\tablenotemark{a}} &
\colhead{Model 3} &
\colhead{Model 4\tablenotemark{a}} &
\colhead{Model 5\tablenotemark{a}} 
}
\startdata
Ion                   &  \ion{O}{7}           &  \ion{O}{7}          &  \ion{Ne}{10}         &  \ion{Ne}{10}        &   \ion{O}{7} \& \ion{Ne}{10} \\
Constant [counts]     & $97.6\pm2.0$          & $97.0^{+1.8}_{-3.9}$ & $78.2^{+1.8}_{-1.8}$ & $78.4^{+1.7}_{-1.7}$  & $97.3^{+1.9}_{-2.1}$ \& $79.9^{+3.2}_{-1.8}$ \\
\sidehead{Absorption system 1}
Center [\kms]         & $-620\pm100$          & $-627^{+78}_{-48}$   & $-543^{+96}_{-59}$    & $-378^{+72}_{-72}$& $-450^{+160}_{-180}$ \\
FWHM [\kms]           & $1250\pm140$          & $560^{+120}_{-180}$  & $1228^{+154}_{-124}$  & $548^{+128}_{-106}$  & $730^{+190}_{-130}$  \\
Normalization [counts]& $-67.3^{+5.9}_{-9.8}$ & $-86^{+17}_{-120}$   & $-53.9^{+3.2}_{-3.4}$ & $-52.3^{+12.2}_{-5.8}$    & $-121^{+119}_{-49}$ \& $-67^{+116}_{-16}$ \\
\sidehead{Absorption system 2}
Center [\kms]         &   \nodata             & $-1284^{+77}_{-38}$  &   \nodata             & $-984^{+118}_{-110}$ & $-1200^{+130}_{-120}$ \\
FWHM [\kms]           &   \nodata             & $386^{+100}_{-74}$   &   \nodata             & $634^{+198}_{-148}$  & $540\pm210$ \\
Normalization [counts]&   \nodata             & $-45.6^{+6.3}_{-39}$ &   \nodata             & $-41.3^{+12.6}_{-5.6}$    & $-36.5^{+18.2}_{-7.1}$ \& $-31.8^{+8.6}_{-14.9}$ \\
\sidehead{Emission} 
Center [\kms]         & $-47\pm35$            & $-140^{+230}_{-620}$ & $283^{+58}_{-53}$    & $371^{+50}_{-52} $  & $-220^{+130}_{-420}$ \\
FWHM [\kms]           & $637^{+100}_{-87}$    & $910^{+1020}_{-350}$ & $541^{+180}_{-135}$ & $275^{+177}_{-120}$&$700^{+420}_{-130}$ \\
Normalization [counts]& $70^{+16}_{-14}$      & $36^{+81}_{-11}$     & $33.4^{+9.8}_{-8.3}$     & $25.1^{+11.6}_{-9.9}$& $103^{+192}_{-45}$ \& $11^{+40}_{-9}$ \\
\sidehead{Statistics} 
Data points           &         58 &    58 &    58 &     58 &  58 \& 58 \\
$\chi^2$              &       68.8 &  44.0 &  44.8 &  36.5  &  109.9  \\
d.o.f.                &         51 &    48 &    51 &     48 &   102    \\
$\chi^2_\nu$          &        1.4 &  0.92 &  0.88 &   0.76 &   1.1    \\
Probability\tablenotemark{b} & 0.049 &  0.64 &  0.72 &  0.89 &  0.28   
\enddata
\vglue-0.6cm
\tablecomments{Uncertainties are 1$\sigma$ (68.3\% confidence)
and were computed using the PROJECTION command in the SHERPA fitting
tool of CIAO. This command varies a given parameter along a grid of
values while the values of all the other free parameters are
allowed to vary to new best-fit values.}
\tablenotetext{a}{Shown in Figure~\protect\ref{linesmodel}.}
\tablenotetext{b}{This is the $Q$-value which measures the probability
that a value of $\chi^2$ as poor as the one found should occur by
chance (Press et al. 1992 \S15.2).}
\end{deluxetable}
\begin{multicols}{2}

\noindent (\ion{Ne}{10}) we compute
an $F$-statistic, $F_\chi = \Delta\chi^2/\chi^2_{\nu 2} = 27.0$ (10.9);
here $\chi^2_{\nu 2}= 0.92$ (0.76) is the reduced $\chi^2$ for the model
with two absorption lines. Since the additional absorption line adds three
parameters the $F_\chi$ per parameter is 9.0 (3.6). The probability of
observing $F_\chi \geq 9.0$ (3.6) for $\nu_1=1$ and $\nu_2=48$ dof is
$\sim 0.5$\% (5\%) (Table C5 of Bevington \& Robinson 1992). Thus, the
confidence level for the presence of two absorption systems instead of one
in \ion{O}{7} is $\sim 99.5$\% and in \ion{Ne}{10} is $\sim 95$\%. Given
that $\chi^2_{\nu 2} < 1$, it is worth examining the significance of
the additional component if we set $\chi^2_{\nu 2}=1$. In that case,
when repeating the above calculation, the confidence level is 99.3\%
for \ion{O}{7} and 90\% for \ion{Ne}{10}. Thus, these data indicate
that there are two X-ray absorption systems detected in \ion{O}{7}
and probably two systems in \ion{Ne}{10}.

The fits presented in Table~\ref{gaussfit} indicate that the outflow
velocities of the two absorption systems in \ion{O}{7} are inconsistent
with those of \ion{Ne}{10} (if two systems exist in the latter).
One possible explanation is that the two ions are present in different
proportion in the different outflowing components, and each has its own
mean outflow velocity. Another explanation is that since the \ion{O}{7}
emission is stronger than the \ion{Ne}{10} emission (as indicated
in Figures~\ref{xrayspec} and \ref{velspec_meg}) the trough of the
\ion{O}{7} absorption is getting filled more than the \ion{Ne}{10}
trough, shifting the \ion{O}{7} trough minimum to higher velocities
than is the case for the \ion{Ne}{10} trough. This latter explanation
is reinforced by comparing the two panels of Figure~\ref{linesmodel}:
while the blue wing of the absorption has approximately the same
velocity structure for the two ions, the red wing of the \ion{O}{7}
absorption is significantly blueshifted when compared to the red wing of
the \ion{Ne}{10} absorption. The stronger emission of \ion{O}{7} seems
to fill more of its absorption trough. We note that Gaussian models for
these lines might not be perfect, and this limitation in the modeling
might cause the \ion{O}{7} and \ion{Ne}{10} absorption systems to have
different apparent velocities due to filling of the trough.

We tried to constrain the latter possibility by fitting the \ion{O}{7}
and \ion{Ne}{10} simultaneously with the same model (note that the
instrumental resolutions nearly the same for both lines as explained
below). We fitted the two velocity spectra with three Gaussians while
constraining the position and the FWHM of each Gaussian to be the
same for each spectrum; the continuum fluxes and the normalizations
of the three Gaussians were left free for each spectrum. The results
are presented as model 5 in Table~\ref{gaussfit} (dashed line in
Figure~\ref{linesmodel}). This model does not fit adequately the data
($\chi^2 = 1.1$); it systematically deviates from the data points near
the absorption troughs, and has a poor fit to the \ion{Ne}{10} emission
feature. Thus, we cannot firmly confirm that the absorption in both ions
arises from the same two systems.

The instrumental resolution, given in Gaussian FWHM, at the wavelengths
of the combined lines is 318--396~\kms\ for the \ion{O}{7} lines
and 353--372 \kms\ for the \ion{Ne}{10} lines. The resolution of the
velocity spectra of Figure~\ref{linesmodel} is a combination of those
resolutions. In the following we consider the worst resolution for
each of the ions, and we use the individual fits for each ion (models
2 and 4 in Table~\ref{gaussfit}). We find that the widths of the two
absorption systems of \ion{O}{7} are consistent with the instrumental
resolution, so they are unresolved individually; upper limits are 550
\kms\ for the low-velocity system and 280 \kms\ for the high-velocity
system. For \ion{Ne}{10}, if two systems exist, the fit (model 4)
seems to resolve the individual systems with FWHMs of $400\pm160$~\kms\
and $510^{+230}_{-200}$~\kms\ for the low and high velocity systems,
respectively. For a single Gaussian model the lines are much broader than
the instrumental width, and their velocity dispersion can be determined,
though these fits are inferior in statistical quality as discussed above.

The above differences between the \ion{O}{7} and the \ion{Ne}{10}
profiles suggest that the absorbers in the two ions have different
dynamical structure, and filling of the troughs by emission only has
a secondary effect. While two distinct narrow absorption systems are
found in \ion{O}{7}, the significance of the two systems in \ion{Ne}{10}
is lower, and both are resolved. This may indicate that we are seeing a
manifestation of a more complex unresolved velocity structure. Such a case
was discussed by Collinge et al. (2001) for NGC\,4051 where as many as 10
UV absorption systems were identified which are consistent in velocity
shift with only one of the two observed X-ray absorption systems. In
the case of NGC\,3783 there are the three UV systems described in \S1
which are marked in Figure~\ref{thirdorder}. While the highest velocity
absorption system in the UV is consistent with the corresponding one seen
in the X-rays, the two lower velocity UV systems cannot be resolved with
the HETGS.\footnote{We note that although the \ion{Ne}{10} Ly$\alpha$
line is clearly resolved by the MEG third-order spectrum (see the bottom
panel of Figure~\ref{thirdorder}), the poor S/N does not enable better
constraints than those presented here.} In addition, while the velocities
of the UV absorption systems agree with the \ion{O}{7} absorbers,
the \ion{Ne}{10} absorption is shifted relative to the UV absorption
systems.  This reinforces the suggestion that while \ion{O}{7} has the
same dynamical structure as the low-ionization lines seen in the UV, the
higher ionization line \ion{Ne}{10} has a different dynamical structure.
This suggestion is also supported by Kraemer et al. (2001) who find
different levels of ionization for different UV absorption systems,
and by the George et al. (1998) result of different X-ray absorbing
components with different ionizations based on variability considerations.

\section{Emission Lines and Radiative Recombination Continua}

The combined 900 ks spectrum shows $\sim 20$ emission lines. These lines
are mainly the He-like triplets (resonance---$r$, intercombination---$i$,
and forbidden---$f$ lines) of \ion{O}{7}, \ion{Ne}{9}, and \ion{Mg}{11}
as well as Ly$\alpha$ lines from the H-like species of these
elements. Measurements of the identified emission lines are presented in
Table~\ref{linetab} (emission lines are noted with negative EWs). The
line velocities are consistent with the systemic velocity of NGC\,3783
with no apparent systematic redshift or blueshift; the average and rms
shift of the 17 emission lines is $130\pm290$~\kms.

The flux ratios of the triplet lines can serve as density and temperature
diagnostics for photoionized media. We tried to apply the theoretical
triplet line ratios calculated by Porquet \& Dubau (2000) in order to
estimate the physical conditions in the emitting gas of NGC\,3783. The
line ratios used in such an analysis are $R(n_{\rm e})=f/i$ and $G(T_{\rm
e})=(f+i)/r$. The calculations account for recombination and collisional
processes. They do not include line pumping (``continuum fluorescence'',
i.e., absorption followed by emission in resonance lines) which depends
strongly on the emission geometry and the line widths. This process can
change the resonance-line intensity in the warm absorber environment
leading to inaccuracy in determining $G(T_{\rm e})$.

\centerline{\includegraphics[width=8.5cm]{f11.eps}}
\figcaption{Emission lines from He-like ion triplets (resonance, 
intercombination, and forbidden lines, as marked in the panels). 
(a) \protect\ion{O}{7} from the MEG binned to 0.01~\AA .
(b) \protect\ion{Ne}{9} from the MEG binned to 0.005~\AA .
(c) \protect\ion{Mg}{11} from the MEG+HEG binned to 0.005~\AA .
The fitted local continuum is plotted as a solid line in each panel.
In (b) and (c) the lines blended with the triplets are marked.
The un-labeled vertical marks in (c) denote the \protect\ion{Ne}{10}
lines which are blended with the \protect\ion{Mg}{11} triplet. All lines
are marked at their rest wavelengths.
\label{triplets} }
\centerline{}
\centerline{}

Using the local continuum (\S\ref{basic_line}) and the line intensities
(Table~\ref{linetab}), we find the measurement of the \ion{Mg}{11} and
\ion{Ne}{9} triplets to be problematic (see Figure~\ref{triplets}). The
\ion{Mg}{11} triplet is heavily blended with \ion{Ne}{10} and iron
lines which makes it unreliable for our analysis. Also, the $r$ line
in \ion{Ne}{9} is blended with iron lines, and this prevents the use
of this line and its resulting $G(T_e)$ in our analysis. However, a
qualitative examination of Figure~\ref{triplets} shows that the sum of the
forbidden and intercombination line intensities are much larger than the
intensity of the recombination line. According to the Porquet \& Dubau
(2000) calculations, $G>4$ (which is consistent with the \ion{Ne}{9}
triplet measurements) implies that the plasma emitting the lines is
photoionization dominated (with little collisional ionization), and the
upper limit on its temperature is $10^6$~K. The density diagnostic for
the \ion{Ne}{9} triplet is $R=5.1\pm2.5$ which constrains the density
to have an upper limit of $2\times10^{11}$~cm$^{-3}$.

The \ion{O}{7} triplet region is free from contaminating lines though the
S/N of the spectrum there is low compared to the shorter wavelengths.
For this ion we find $G=3.0\pm1.7$ and $R=2.9\pm1.7$. Such a $G$ may
indicate the plasma is not purely photoionized and collisional effects
are important. It can also indicate that radiation pumping is significant
since it preferentially adds to the intensity of the $r$ line resulting
in small value of $G$. A proper modeling which accounts for all the
above effects will be presented in H. Netzer et al. (in preparation). As
the errors on the fluxes of the \ion{O}{7} $i$ and $f$ lines are large
(Table~\ref{linetab}), we are only able to use the density diagnostic $R$
to constrain the density to have an upper limit of $10^{11}$~cm$^{-3}$.

Two emission features at 16.695~\AA\ and 22.376~\AA\ are identified near
the \ion{O}{7} and \ion{N}{6} edges, respectively. We identify these
as radiative recombination continuum (RRC) emission. The widths of
these features can be used to estimate the temperature of the plasma.
The RRC emission feature should peak at the edge energy and sharply
decrease toward higher energies, with a temperature-dependence profile.
The highest velocity shift we find in the \ion{O}{7} RRC emission
is $\sim$2000~\kms. This can be used to estimate a lower limit of
$\sim$60,000~K on the temperature of the plasma.\footnote{We used the
formalism as in Rybicki \& Lightman (1979) with radiative recombination
coefficients from Aldrovandi \& Pequignot (1973).} The \ion{N}{6} RRC
emission yields the same result.

\section{The Iron K${\alpha}$ Line Region}
\label{ironsec}

Kaspi et al. (2001) detected a narrow Fe\,K$\alpha$ line around 6.4
keV in the high-resolution X-ray spectrum of NGC\,3783 but were only
able to place an upper limit of 3250~\kms\ on its FWHM due to the
limited S/N. Our combined 900 ks exposure has allowed us to measure
this line precisely. Figure~\ref{fe_plot} shows the HEG spectrum in the
region of the Fe\,K$\alpha$ line. A Gaussian fitted to the 0.0025~\AA\
binned spectrum (Figure~\ref{fe_plot}b) gives a central wavelength of
$1.9378\pm0.0010$~\AA\ ($6398.2\pm3.3$~eV) which is consistent with the
Fe\,K$\alpha$ line from \ion{Fe}{1} to \ion{Fe}{11} (at the rest frame
velocity system of NGC\,3783). Interestingly, the Fe\,K$\alpha$ line
in Figure~\ref{fe_plot}b shows two peaks (though these are {\em not}
resolved) which are consistent with the two expected Fe\,K$\alpha$
lines for \ion{Fe}{1}, K$\alpha_1$ at 1.936~\AA\ (6403.84 eV) and
K$\alpha_2$ at 1.940~\AA\ (6390.84 eV) and a branching ratio of 2:1
(Bearden 1967; Bambynek et al. 1972). We fitted the HEG spectrum with
two Gaussians fixed at the wavelengths of the Fe\,K$\alpha$ lines and
with the same branching ratio. We find the FWHM of the Gaussians to be
$16.3^{+1.7}_{-1.5}$~m\AA\ and, when taking into account the instrumental
FWHM of 12~m\AA, we get a true FWHM of $11.1\pm 2.3$~m\AA. This FWHM
corresponds to $1720\pm360$~\kms\ at the wavelength of the Fe\,K$\alpha$
line. Fitting the data with only one Gaussian yields a consistent result
of FWHM$_{\rm true} = 1860\pm340$~\kms.

The gas emitting the optical emission lines in NGC\,3783 shows an
increase in its density, ionization parameter, and velocity dispersion
of its emitting clouds toward the central ionizing source (Pelat,
Alloin, \& Fosbury 1981; Atwood, Baldwin, \& Carswell 1982; Evans
1988; Winge et al. 1992). The FWHM velocity dispersion in the broad
line region (BLR) of NGC\,3783 is $\sim$ 4000 km\,s$^{-1}$ (Reichert
et al. 1994; Wandel, Peterson, \& Malkan 1999; FWHM$_{\rm H\beta}
= 4100\pm 1160$ \kms). The FWHM velocity dispersion in the narrow
line region (NLR) is $\sim$\,100--700~\kms\ (Pelat et al. 1981; Evans
1988). Winge et al. (1992) analyzed the optical spectrum of NGC\,3783,
decomposing the broad and narrow emission lines into several Gaussians
components. They modeled the optical spectrum with four main regions:
narrow (FWHM$\leq$835~\kms), intermediate (835$<$FWHM$<$1670 \kms), broad
(3000$<$FWHM$<$3340~\kms), and very broad (FWHM$\geq$6680~\kms). The FWHM
we find for the narrow Fe\,K$\alpha$ line ($\sim 1700$~\kms) falls at the
edge of their intermediate region toward the broad region.  Assuming a
simple anticorrelation of line width with radial location in a virialized
system, the origin of this feature could be between the BLR and the 
%
\centerline{\includegraphics[width=8.5cm]{f12.eps}}
\figcaption{HEG spectrum around the Fe\,K$\alpha$ feature: (a) binned to
0.005~\AA\ and (b) binned to 0.0025~\AA. In (b) the two Fe\,K$\alpha$ lines
are marked. Also shown are the theoretical wavelengths of \ion{Fe}{25}
and \ion{Fe}{26} absorption lines as well as the forbidden and
intercombination lines of \ion{Fe}{25} (dotted line).
\label{fe_plot} }
\centerline{}
\centerline{}
\noindent NLR,
which is the location of the putative torus.  In a recent study Onken
\&  Peterson (2002), find the BLR in NGC\,3783 is virialized and deduce
the central mass to be $(8.7 \pm 1.1)\times 10^6M_\sun$. Adopting this
central mass and using equation 5 of Kaspi et al. (2000b), we estimate
that the narrow Fe\,K$\alpha$ line is emitted at a distance of $\sim
20$ lt-days ($\sim 5.2\times10^{16}$ cm) from the central source. This
places the narrow Fe\,K$\alpha$ region at the outer parts of the BLR
or the inner part of the torus. This result is similar to the one found
found by Yaqoob et al. (2001) for the narrow Fe\,K$\alpha$ emission line
in NGC\,5548. Detailed reverberation mapping studies may be able to test
this result. The EW of the narrow Fe\,K$\alpha$ line is $27.4\pm3.3$~m\AA\
($90\pm11$~eV), and its flux is $(5.26\pm0.63)\times10^{-5}\ {\rm photons\
cm^{-2}\ s^{-1}}$. This is consistent with the predicted flux arising
from a central torus (see the discussion in \S7.1 of Kaspi et al. 2001).

We searched the spectrum for narrow neutral lines from abundant elements
other than iron and found none. A standard reflection model predicts
such lines to have EWs of less than about one eV (e.g., Matt, Fabian,
\& Reynolds 1997) which cannot be detected in our data.  We also do not
detect neutral iron K edge at 7.1 keV. The derived upper limit for the
absorption optical depth of such an edge is 0.1.

Looking at the binned spectrum in Figure~\ref{xrayspec}, it is clear
that the narrow Fe\,K$\alpha$ line has a more complex profile than that
of a single Gaussian and seems to have a red wing extending to $\sim
2$~\AA\ ($\approx 6.2$~keV). A plausible explanation for this red wing
is the ``Compton shoulder'' (e.g., Iwasawa, Fabian, \& Matt 1997 and
references therein) produced by Compton scattering in Compton thick,
cold material like the one suggested for the obscuring torus. The
6.2--6.4 keV shoulder has a total flux of $(8.6\pm2.7)\times10^{-6}\
{\rm photons\ cm^{-2}\ s^{-1}}$, and an EW of $4.2\pm1.3$ m\AA, i.e.,
$\sim 14$\% of the total line flux. These numbers are in agreement
with previous models and observations of such a shoulder (e.g.,
Iwasawa et al. 1997), and indicate that the line comes from matter
that is Compton thick.

The X-ray spectrum (Figure~\ref{fe_plot}) shows a hint of the
Fe\,K$\beta$ line expected to be at 1.757~\AA\ (7058 eV) and predicted
to have about 11\% of the Fe\,K$\alpha$ flux. Figure~\ref{fe_plot}
also shows the line features expected from \ion{Fe}{25} and
\ion{Fe}{26}. There are possibly two absorption lines identified from
\ion{Fe}{25}, at 1.573 \AA\ (EW=$6.6\pm3.8$~m\AA) and 1.850 \AA\
(EW=$3.8\pm1.4$~m\AA). However, due to the poor S/N and resolution
($R\approx130$), we can make no strong claim about the reality of
these features.

A broad Fe\,K$\alpha$ line has been observed in {\it ASCA} spectrum
of NGC\,3783 (Nandra et al. 1997; George et al. 1998), attributed
to emission from the inner accretion disk (Fabian et al. 1989). The
superior resolution of the {\it Chandra} gratings can be used to
deconvolve any narrow components of the line, which can significantly
affect the derived spectral parameters (e.g., Weaver \& Reynolds 1998;
Yaqoob et al. 2001). In order to explore whether these {\it Chandra}
data indicate the presence of a broad, disk-line component requires a
global model that accounts fully for the effects of ionized absorption
on the continuum. A full treatment of the broad Fe\,K$\alpha$ line will
be presented in a future paper.

\section{Summary}

The bright Seyfert~1 galaxy NGC\,3783 was observed with the {\em
Chandra}/HETGS for a total of $\sim$850~ks during the Spring of
2001 as part of an intensive study which also included {\em RXTE},
{\em HST}/STIS, {\em FUSE}, and ground-based observations. In this
paper we have presented the mean 900~ks high-resolution X-ray spectrum
(including 56~ks of observation from January 2000). 
Our main results are as follows:

\begin{enumerate}

\item The 900 ks spectrum reveals strong absorption lines from
H-like and He-like ions of N, O, Ne, Mg, Al, Si, and S. We also
detect inner-shell lines from lower ionization ions of Mg, Si, and
S. We suggest a possible detection of lines from H-like and He-like
ions of Ar and Ca, and from H-like C. There are also many absorption
lines from iron ions; L-shell lines of \ion{Fe}{17}--\ion{Fe}{24},
lines from Fe M-shell ions (UTA), and possibly resonance lines from
\ion{Fe}{25} are seen.

\item Out of the 135 detected absorption features we consider 42
to be unblended and to have good S/N. The mean outflow velocity
obtained from these lines is $-590\pm150$~\kms, and the mean FWHM
is $820\pm280$~\kms. We do not find any correlation of the velocity
shifts or the FWHMs with ionization potentials. We further do not
find these characteristics to be consistent with a constant value,
probably due to line emission ``filling in'' the absorption.

\item We resolve the \ion{O}{7} lines into two outflowing systems
at $-627^{+78}_{-48}$ and $-1284^{+77}_{-38}$~\kms\ which are not
individually resolved. The detection of the second \ion{O}{7}
absorption system confirms its prediction by Kraemer et
al. (2001) which was based on their model for the UV absorption
systems. The \ion{Ne}{10} lines possibly have two absorption systems at
$-378^{+72}_{-72}$ and $-984^{+118}_{-110}$~\kms\ which are marginally
resolved to have FWHMs of $\sim 600$~\kms. The detection of these
systems is only marginally statistically significant. We do not find
the \ion{O}{7} velocity structure to be entirely consistent with the
structure of \ion{Ne}{10}, suggesting the two ions are arising from
regions that are somewhat dynamically distinct.

\item The outflow velocities we find for the two X-ray absorbing
systems are consistent with the three absorption systems found
in the UV at about $-1400$, $-720$, and $-$560 \kms\ (Kraemer et
al. 2001). Although the HETGS does not have the resolution to resolve
these three systems, the overall velocity range is the same; thus it
seems plausible that there are more than two absorption systems in
the X-ray spectrum.

\item We have attempted a detailed curve of growth analysis of the
spectrum. However, three factors limit such an analysis: (1) The
probable presence of unresolved multiple absorption systems must
be taken into account, (2) In several cases it appears that the
emission line adjacent to the blueshifted absorption is filling
the absorption. Thus, the measured EWs do not represent the true
absorption, and (3) It is likely that some absorption lines are
saturated though they are not black due to the limited resolution,
the filling of the trough with emission, and scattered X-ray
continuum photons. In our curve of growth analysis we used the
simplified assumption of having one absorption system with Doppler
velocity of 350~\kms and list the results in Table~4. In
particular we find the predicted absorption edge optical depths
of \ion{O}{7} and \ion{O}{8} to be $0.26^{+0.15}_{-0.10}$ and
$0.42^{+0.33}_{-0.19}$ which indicate that oxygen edges {\it are
present} in spectrum. Furthermore, these values are too small to
explain the observed absorption-edge optical depths. To explain this
one needs to assume at least two absorption systems with a Doppler
velocities of 120~\kms. This yields $\tau_{\rm O\,VII} (739~{\rm
eV}) = 0.4^{+0.4}_{-0.2}$ and $\tau_{\rm O\,VIII} (871~{\rm eV})
= 1.3^{+2.9}_{-0.8}$.

\item  About two-dozen emission features are detected in the spectrum
mainly from the He-like triplets of \ion{O}{7}, \ion{Ne}{9}, and
\ion{Mg}{11} as well as the Ly$\alpha$ lines from the H-like species
of these elements.  The emission lines are consistent with being at
the systemic velocity of NGC\,3783.

\item  We detect radiative recombination continuum emission near the
edges of \ion{N}{6} and \ion{O}{7}. The lower limit on the temperature
derived from the width of these RRCs is $\sim 60,000$~K.

\item  Table~\ref{linetab} and Figure~\ref{linesmodel} show that
for almost all lines the emission EW is smaller than the absorption
EW. This indicates that the global covering factor must be smaller
than the line-of-sight covering factor (which is $\sim~1$). This is
consistent with the covering factors used in the model presented by
Kaspi et al. (2001) and will be discussed in a future publication.

\item   We resolve the narrow Fe\,K$\alpha$ line to have a FWHM
of $1720\pm360$~\kms. This line is at $1.9378\pm0.0010$~\AA\
($6398.2\pm3.3$~eV) and has an EW of $27.4\pm3.3$~m\AA\
($90\pm11$~eV). We detect a small (EW = $4.2\pm1.3$ m\AA) red wing
of the line which extends down to 6.2 keV. A likely explanation for
the red wing is a ``Compton shoulder'' which indicates the presence
of cold, Compton-thick gas.  The treatment of the broad component
of the Fe\,K$\alpha$ line (which has been observed in this object
in previous {\it ASCA\/} observations) requires a global model
that accounts fully for the effects of ionized absorption on the
continuum. This will be discussed in a future paper.

\end{enumerate}

In this paper we have presented the best mean spectrum (in terms of
the combination of signal-to-noise and resolution) ever obtained for an
AGN in the X-ray band. We presented the measured properties of the many
line features detected in the spectrum and simple interpretations. An
accompanying paper (I. M. George et al., in preparation) will present
the time variability properties of this spectrum. Subsequent work will
focus on modeling this superb spectrum and on its relation to other
wavelength bands. This high-resolution spectrum demonstrates the wealth
of information that can be extracted from high resolution X-ray data.
Combined with information from all other wavelength bands, it is leading
to a much deeper understanding of the inner regions of AGNs.

\acknowledgments

We thank the anonymous referee for constructive comments.
We gratefully acknowledge the financial support of CXC grant GO1-2103
(S. K., W. N. B., I. M. G.), NASA LTSA grant NAG~5-8107 (S. K., W. N.
B.), the Alfred P. Sloan Foundation (W. N. B.), an ISF grant (H. N., S. K.),
and NSF grant AST-9984040 (F. W. H.). This work would not have been
possible without the enormous efforts of the entire {\it Chandra} team.

\end{document}